\documentclass{aa}

\usepackage[english]{babel}
\usepackage{graphicx}
\usepackage[varg]{txfonts}
\usepackage{natbib}
\usepackage{longtable}
\usepackage{array}
\usepackage{multirow}
\usepackage{amsmath}

\usepackage{setspace}

\bibpunct{(}{)}{;}{a}{}{,}
\begin{document}

\title{Laboratory rotational spectroscopy of acrylamide and search for
acrylamide and propionamide toward Sgr B2(N) with ALMA\thanks{Tables A.1. and A.4. are only available in electronic form
at the CDS via anonymous ftp to cdsarc.u-strasbg.fr (130.79.128.5) or via http://cdsweb.u-strasbg.fr/cgi-bin/qcat?J/A+A/
}}

   \author{L. Kolesnikov\'{a}\inst{1}
   \and A. Belloche\inst{2}
   \and J. Kouck\'{y}\inst{1}
   \and E. R. Alonso\inst{3}
   \and R. T. Garrod\inst{4}
   \and K. Lukov\'{a}\inst{1}
   \and K. M. Menten\inst{2}
   \and H.~S.~P.~M\"uller\inst{5}
   \and P. Kania\inst{1}
   \and \v{S}. Urban\inst{1}
          }

\institute{Department of Analytical Chemistry, University of Chemistry and Technology,
Technick\'{a} 5, 166 28 Prague 6, Czech Republic\\\email{lucie.kolesnikova@vscht.cz}
\and
Max-Planck-Institut f\"{u}r Radioastronomie, Auf dem H\"{u}gel 69, 53121 Bonn, Germany
\and Grupo de Espectroscopia Molecular (GEM), Edificio Quifima, \'{A}rea de Qu\'{\i}mica-F\'{\i}sica, Laboratorios de Espectroscopia y Bioespectroscopia, Parque Cient\'{\i}fico UVa, Unidad Asociada CSIC, Universidad de Valladolid, 47011 Valladolid, Spain
\and
Departments of Chemistry and Astronomy, University of Virginia, Charlottesville, VA 22904, USA
\and
Astrophysik/I. Physikalisches Institut, Universit{\"a}t zu K{\"o}ln, Z{\"u}lpicher Str. 77, 50937 Cologne, Germany}

\date{Received ; accepted }

\titlerunning{The rotational spectrum and ISM search of acrylamide}
\authorrunning{Kolesnikov\'{a} et al.}


\abstract
   {Numerous complex organic molecules have been detected in the universe among which amides are considered as models for species containing the peptide linkage. Acrylamide (CH$_{2}$CHC(O)NH$_{2}$) bears in its backbone not only the peptide bond, but also the vinyl functional group which is a common motif in many interstellar compounds. This makes acrylamide an interesting candidate for a search in the interstellar medium. In addition, a tentative detection of the related molecule propionamide (C$_{2}$H$_{5}$C(O)NH$_{2}$) has been recently claimed toward Sgr B2(N).}
   {The aim of this work is to extend the knowledge of the laboratory rotational spectrum of acrylamide to higher frequencies, which makes possible to conduct a rigorous search for interstellar signatures of this amide by millimeter wave astronomy.
      }
   {The rotational spectrum of acrylamide was investigated between 75 and 480 GHz. After its detailed analysis, we searched for emission of acrylamide in the imaging spectral line survey ReMoCA performed with the Atacama Large Millimeter/submillimeter Array toward Sgr B2(N). We also searched for propionamide in the same source. The astronomical spectra were analyzed under the assumption of local thermodynamic equilibrium.}
   {We report accurate laboratory measurements and analyses of thousands rotational transitions in the ground state and two excited vibrational states of the most stable \textit{syn} form of acrylamide. In addition, we report an extensive set of rotational transitions for the less stable \textit{skew} conformer. Tunneling through a low energy barrier between two symmetrically equivalent configurations has been revealed for this higher-energy species. 
   Neither acrylamide nor propionamide were detected toward the two main hot
   molecular cores of Sgr~B2(N). We did not detect propionamide either toward
   a position located to the east of the main hot core, thereby not confirming
   the recent claim of its interstellar detection toward this position. We find
   that acrylamide and propionamide are at least 26 and 14 times, respectively,
   less abundant than acetamide toward the main hot core Sgr~B2(N1S), and at
   least 6 and 3 times, respectively, less abundant than acetamide toward the
   secondary hot core Sgr~B2(N2).}
   {A comparison with results of astrochemical kinetics model for related
   species suggests that acrylamide may be a few hundred times less abundant
   than acetamide, corresponding to a value at least an order of magnitude
   lower than the observational upper limits. Propionamide may be as little
   as only a factor of two less abundant than the upper limit derived toward
   Sgr~B2(N1S). Last but not least, the spectroscopic data presented in this work will aid future searches of acrylamide in space.}


   \maketitle
%

\section{Introduction}
\label{sect_intro}

The peptide bond, --C(=O)NH--, found in amides is vital for biology. It provides linkages between amino acids, giving rise to peptides, which are the key players in mediating both the molecular interactions underpinning present life on Earth and potentially the prebiotic processes that preceded it \citep{Frenkel-Pinter2020,Weber2006}. Therefore, it is not surprising that the question of peptide bond formation represents a highly topical theme in different fields and disciplines, including prebiotic astrochemistry \citep{Ruiz-Mirazo2014,Sandford2020,Stolar2021}.

In astronomical settings, amides and other peptide-like species are actively searched for in star-forming regions, solar-type protostars and other environments \citep[e.g.][]{Halfen2011,Belloche2017,Mendoza2014,Ligterink2020,Ligterink2021,Gorai2020,Colzi2021}.
These studies uncover their formation in space and their role as nodes in chemical networks connecting other prebiotic molecules.
Remarkable progress has been made in increasing the sensitivity of astronomical observing capabilities \citep{Jorgensen2020,Tercero2021}, opening up unique possibilities to observe new target compounds in lower abundances than before. Much experimental work has concentrated on laboratory simulations designed to mimic the interstellar ices and provided compelling evidence that molecules containing the peptide bond can be formed abiotically \citep[see, e.g.,][]{Kaiser2013,Ligterink2018,Frigge2018}. Finally, possible formation and destruction pathways of peptide-like molecules became an important target in theoretical calculations and chemical modeling \citep[see, e.g.,][]{Redondo2013,Redondo2014,Rimola2014,Barone2015,Quenard2018}. All these studies provide valuable information on the peptide bond chemistry in space from different viewpoints, but sharing the same goal: to contribute to the deciphering of one of the most enduring puzzles presented to humankind: the origins of life.

Laboratory rotational spectroscopy is one of the key pillars in decoding the prebiotic inventory of space. It has contributed to the detection of several amides
among other prebiotically interesting molecules \citep[][]{McGuire2018} of which
glycolaldehyde \citep[CH$_{2}$OHCHO;][]{Hollis_2000}, aminoacetonitrile \citep[NH$_{2}$CH$_{2}$CN;][]{Belloche2008}, hydroxylamine \citep[NH$_2$OH;][]{Rivilla2020}, and ethanolamine \citep[NH$_2$CH$_2$CH$_2$OH;][]{Rivilla2021} are mentioned as examples.
The simplest member of the amide family, formamide (NH$_2$CHO), has been observed in spectral-line rich sources, namely Sgr~B2 \citep{Rubin1971,Gottlieb1973,Nummelin1998,Turner1989,Thiel2017}, Orion~KL \citep{Turner1989,Motiyenko2012}, IRAS~16293–2422 \citep{Kahane2013,Coutens2016} and, more recently, in many other sources \citep[see, e.g.,][]{Lopez-Sepulcre2019}. Also recently, N-methylformamide (CH${_3}$NHCHO) has been discovered toward the Sgr B2 and NGC 6334I star-forming regions \citep{Belloche2017,Belloche2019,Ligterink2020} in which its structural isomer acetamide (CH$_3$C(O)NH$_2$) had also been detected \citep{Hollis2006,Halfen2011,Belloche2013,Ligterink2020}. Very recently, a tentative detection of propionamide (CH$_{3}$CH$_{2}$C(O)NH$_{2}$) toward Sgr B2 has been claimed by \cite{Li2021}. These detections suggest that more peptide bond bearing molecules are yet to be detected.

Another species with the peptide link is acrylamide (CH$_{2}$CHC(O)NH$_{2}$). It contains two hydrogen atoms fewer than propionamide and shares the vinyl functional group with other known interstellar compounds such as vinyl cyanide \citep[CH$_{2}$CHCN;][]{Gardner1975}, vinyl alcohol \citep[CH$_{2}$CHOH;][]{Turner2001}, propenal  \citep[CH$_{2}$CHCHO;][]{Hollis2004}, propylene \citep[CH$_{2}$CHCH$_{3}$;][]{Marcelino2007}, vinylacetylene \citep[CH$_{2}$CHCCH;][]{Cernicharo2021c}, and vinylcyanoacetylene \citep[CH$_{2}$CHC$_{3}$N;][]{Kelvin_Lee_2021}.

Acrylamide is a complex organic molecule (COM) in the astronomical sense (i.e., containing six atoms or more; \citealt{Herbst2009}).
Since COMs might be present in multiple conformations which, in addition, might be governed with large amplitude motions and low-frequency vibrations, their dense rotational spectra might be difficult to interpret. Extensive spectroscopic studies of complex amides such as
N-methylformamide \citep{Belloche2017}, glycolamide \citep{Sanz-Novo2020}, propiolamide \citep{Alonso2021}, and glycinamide \citep{Kisiel2021} were thus conducted to enable their interstellar hunt in the millimeter wave region. However, so far this has not been the case for acrylamide for which the rotational spectrum has been studied only up to 60 GHz \citep{Marstokk2000}. This makes predictions at higher frequencies rather uncertain impeding its rigorous search in space.

      \begin{figure}[ht]
   \centering
   \includegraphics[trim = 0mm 0mm 0mm 0mm, clip, width=8.7cm]{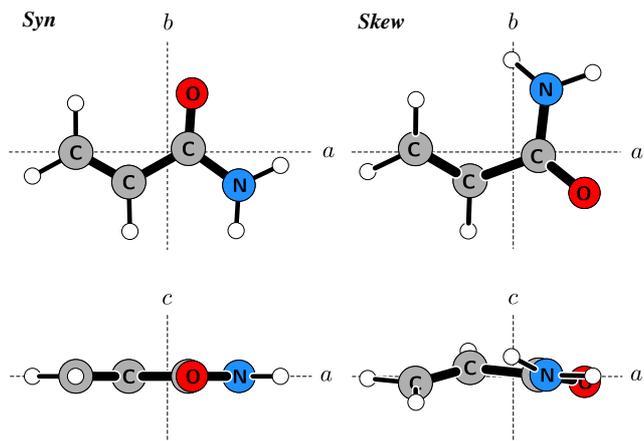}
      \caption{The two conformers of acrylamide depicted in their principal $ab$ and $ac$ inertial planes.
      }
      \label{conformers}
   \end{figure}

As an isolated molecule in the gas phase, acrylamide exists in the form of \textit{syn} and \textit{skew} conformers (see Fig. \ref{conformers}), of which the latter was found by \cite{Marstokk2000} to be 6.5(6)~kJ~mol$^{-1}$ or 543(50)~cm$^{-1}$ less stable. In course of the present work, we measured the rotational spectrum of acrylamide between 75 and 480~GHz and analyzed the spectral signatures of both species. The new measurements and analyses presented here provided precise laboratory information to search for this amide in the high-mass star-forming region Sgr B2(N), a possible source of large peptide molecules.

The remainder of the paper is structured as follows.
Details about the laboratory spectroscopy are given in Sections ~\ref{s:experiments} and ~\ref{s:analysis}
while the search for acrylamide toward Sgr B2(N) is reported in Sect. ~\ref{s:astro}.
In the same Section, we also report our search for propionamide in Sgr B2(N) which
does not confirm its recently claimed interstellar detection.
We discuss our results in Sect.~\ref{s:discussion} and
provide our conclusions in Sect.~\ref{s:conclusions}.


\section{Experiments}
{\label{s:experiments}}

White crystalline acrylamide (m.p. 82--86 $^{\circ}$C) was obtained commercially and was used without any further purification.
The room-temperature rotational spectrum was measured in the frequency region 75--480 GHz using two spectrometers.
The lowest-frequency (75--110 GHz) and the highest-frequency (170--480 GHz) sections were recorded using the millimeter wave spectrometer at the University of Valladolid \citep{Daly2014} which is based on sequential multiplication of an Agilent synthesizer produced frequency by a set of active and passive multipliers from Virginia Diodes, Inc.
The absorption free-space cell was a glass tube (10 cm diameter, 360 cm long). The Prague millimeter wave spectrometer \citep{Kania2006} which is based on the same principle was used to measure the rotational spectrum between 128 and 170 GHz. Here, the free-space cell was 280 cm long with a diameter of 8 cm.
The optical path length was doubled by a roof-top mirror in both cases.
The synthesizer output was frequency modulated in both spectrometers (modulation frequency of 10.2 and 28 kHz) and the detection system was completed by a demodulation procedure achieved by a lock-in amplifier that was tuned to twice the modulation frequency. All spectra were registered with the sample pressure between 10 and 20 $\mu$bar by upward and downward frequency scanning and averaged. The individual spectral sections were merged into a single spectrum and subjected to the analysis using the Assignment and Analysis of Broadband Spectra (AABS) package \citep{aabs,Kisiel2012}.

      \begin{figure*}[ht]
   \centering
   \includegraphics[trim = 40mm 20mm 95mm 0mm, clip, width=17.0cm]{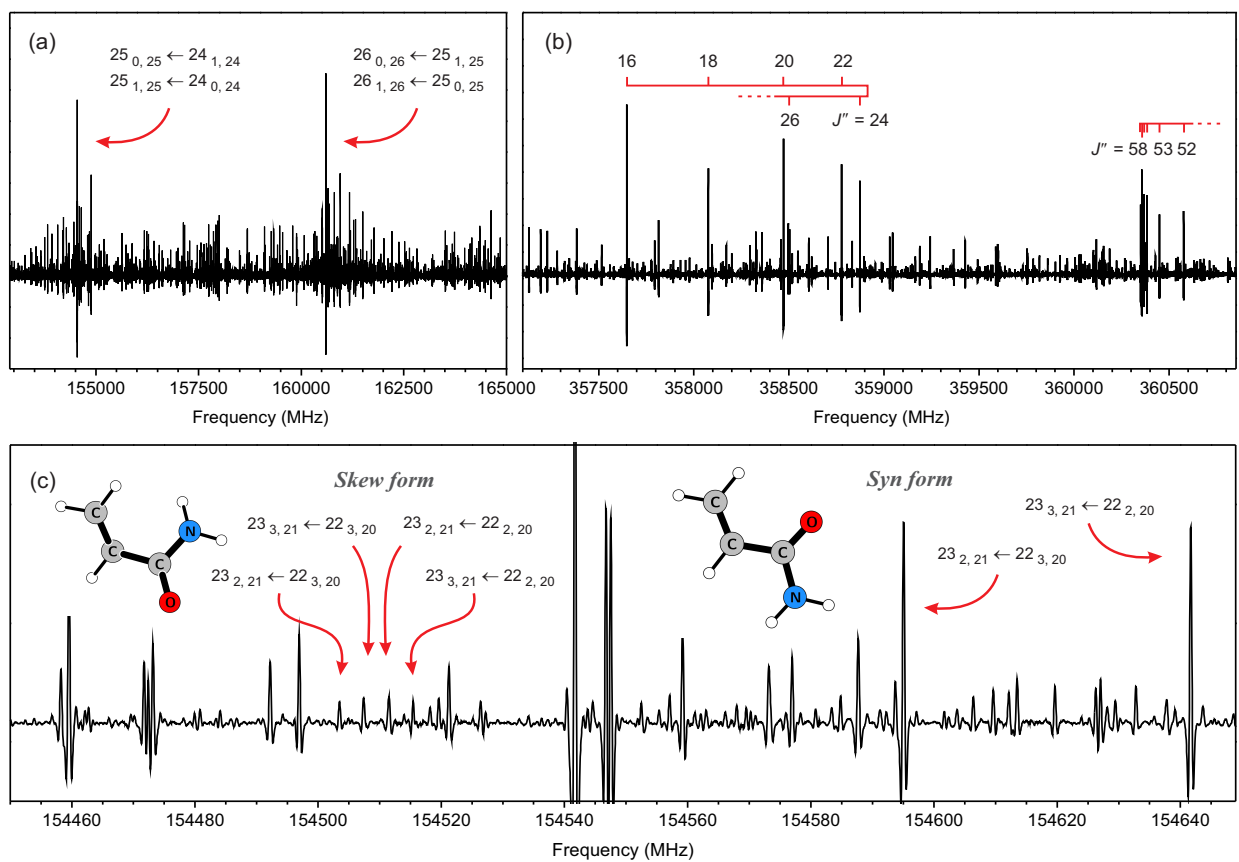}
      \caption{Main features in the room-temperature rotational spectrum of acrylamide. (a) Strong lines of the \textit{syn} conformer in the millimeter wave region consisting of a pair of $b$-type transitions involving $K_{a}$ = 0, 1 and $K_{c} = J$ energy levels. (b) Groups of $b$-type R-branch transitions near the sub-millimeter wave region. A diffuse group on the left embraces doubly degenerate transitions with the same $K_{a}$. Here, the leading transition is $J = K_{a} = 17 \leftarrow 16$ and the value of $J$ increases by 2 while the value of $K_{a}$ decreases by 1 with each successive line. The compact group on the right encompasses degenerate transitions with two different $K_{a}$. In the present group, degenerate $K_{a}= 0, 1$, $K_{a}= 1, 2$, ... $K_{a}= 6, 7$ transitions are observed. Only $J$ quantum numbers are labeled with the red combs for simplicity. (c) A portion of the spectrum showing the rotational transitions of \textit{syn} and \textit{skew} acrylamide that involve $22_{~2,~20}$ and $22_{~3,~20}$ energy levels. While two $a$-type and two $b$-type transitions are observed for the \textit{skew} conformer, only two $b$-type transitions are found for the \textit{syn} form in agreement with their dipole moment components.}
      \label{features}
   \end{figure*}

\begin{table*}[ht]
\caption{Spectroscopic constants of \textit{syn} acrylamide in the ground state (G.S.) and two excited vibrational states ($A$-reduction, I$^{\text{r}}$-representation).}
\label{T1}
\begin{center}
\begin{footnotesize}
\setlength{\tabcolsep}{4.0pt}
\begin{tabular}{ l r r r r r r r }
\hline\hline
 & \multicolumn{3}{c}{This work} &  & \multicolumn{3}{c}{\cite{Marstokk2000}}  \\
\cline{2-4}
\cline{6-8}
 & G.S.  & $v_{\text{24}}=1$  & $v_{\text{24}}=2$ &  & G.S. & $v_{\text{24}}=1$ & $v_{\text{24}}=2$ \\
\hline
$A                   $  /               MHz     &   10732.819331 (65)\tablefootmark{a} &   10660.711630 (73)  &   10594.14406 (10)            & &  10732.8296 (34)  &  10660.7092 (33) &  10594.1266 (38)  \\
$B                   $  /               MHz     &    4218.690256 (41)                  &    4216.462772 (50)  &    4214.275015 (87)           & &   4218.7012 (13)  &   4216.4693 (11) &   4214.2765 (14)  \\
$C                   $  /               MHz     &    3030.752979 (33)                  &    3038.064954 (36)  &    3044.891587 (53)           & &   3030.7434 (13)  &   3038.0502 (17) &   3044.8733 (16)  \\
$\Delta_{J}          $  /               kHz     &       0.786816 (13)                  &       0.804212 (16)  &       0.823406 (30)           & &      0.7043 (33)  &     0.696 (14)   &      0.776 (7)    \\
$\Delta_{JK}         $  /               kHz     &       3.755014 (41)                  &       3.817382 (52)  &       3.91194 (12)            & &      3.370 (22)   &     3.15 (7)     &      3.658 (31)   \\
$\Delta_{K}          $  /               kHz     &       5.49540 (11)                   &       5.17392 (12)   &       5.10038 (19)            & &      5.403 (8)    &      5.100 (24)  &      5.055 (15)   \\
$\delta_{J}          $  /               kHz     &       0.2227676 (56)                 &       0.2240435 (76) &       0.224728 (17)           & &      0.2417 (7)   &      0.2394 (26) &      0.2418 (19)  \\
$\delta_{K}          $  /               kHz     &       3.361833 (72)                  &       3.33546 (10)   &       3.36995 (27)            & &      3.20 (4)     &     3.37 (7)     &      3.41 (5)     \\
$\Phi_{J}            $  /               mHz     &       0.1763 (16)                    &       0.2180 (22)    &       0.4452 (56)             & &     --66 (6)      &     --208 (33)   &     --7 (7)       \\
$\Phi_{JK}           $  /               mHz     &       1.361 (16)                     &    --0.632 (24)      &       1.361\tablefootmark{b}  & &    --222 (34)     &    --2300 (400)  &    --650 (110)    \\
$\Phi_{KJ}           $  /               mHz     &    --30.836 (56)                     &   --34.835 (86)      &    --47.69 (10)               & &    --1540 (130)   &    --3500 (500)  &    --570 (170)    \\
$\Phi_{K}            $  /               mHz     &      55.087 (76)                     &      44.373 (96)     &      62.03 (14)               & &     570 (40)      &     380 (40)     &     110 (100)     \\
$\phi_{J}            $  /               mHz     &       0.09118 (77)                   &       0.0947 (11)    &       0.2032 (30)             & &    1.70 (34)      &    --2.2 (4)     &    --10.7 (26)    \\
$\phi_{JK}           $  /               mHz     &       1.104 (16)                     &       1.065 (23)     &       3.558 (56)              & &    --600 (50)     &    --740 (120)   &    640 (130)      \\
$\phi_{K}            $  /               mHz     &      40.43 (11)                      &      17.00 (18)      &      31.43 (27)               & &     1980 (160)    &    --1850 (170)  &    400 (340)      \\
$J_{\text{min}}/J_{\text{max}}$                 &    3 / 78                            &     4 / 78           &     2 / 78                    & &    1 / 75         &    1 / 68        &   2 / 64          \\
$K_{a}^{\text{min}}/K_{a}^{\text{max}}$         &    0 / 37                            &     0 / 38           &     0 / 37                    & &    0 / 37         &   0 / 40         &   0 / 39          \\
$N$\tablefootmark{c}                            &    2710                              &     2242             &     1599                      & &   354             &   322            &   317             \\
$\sigma_{\text{fit}}$\tablefootmark{d}/ MHz     &    0.028                             &     0.024            &     0.035                     & &   0.081           &   0.063          &   0.090           \\
$\sigma_{\text{w}}$\tablefootmark{e}            &    0.93                              &     0.98             &     0.98                      & &    ...            &   ...            &   ...             \\
\hline
\end{tabular}
\end{footnotesize}
\end{center}
\tablefoot{
\tablefoottext{a}{The numbers in parentheses are the parameter uncertainties in units of the last decimal digit. Their values are close to 1$\sigma$ standard uncertainties (67\% confidence level) because the unitless (weighted) deviation of the fit is close to 1.0. SPFIT/SPCAT program package \citep{Pickett1991} was used for the analysis.}
\tablefoottext{b}{Fixed to the ground state value which is usually a preferred constraint against the zero or poorly determined value \citep{Urban1990,Koucky2013}.}
\tablefoottext{c}{Number of distinct frequency lines in the fit.}
\tablefoottext{d}{Root mean square deviation of the fit.}
\tablefoottext{e}{Unitless (weighted) deviation of the fit.}}
\end{table*}

      \begin{figure}
   \centering
   \includegraphics[trim = 22mm 20mm 25mm 16mm, clip, width=8.1cm]{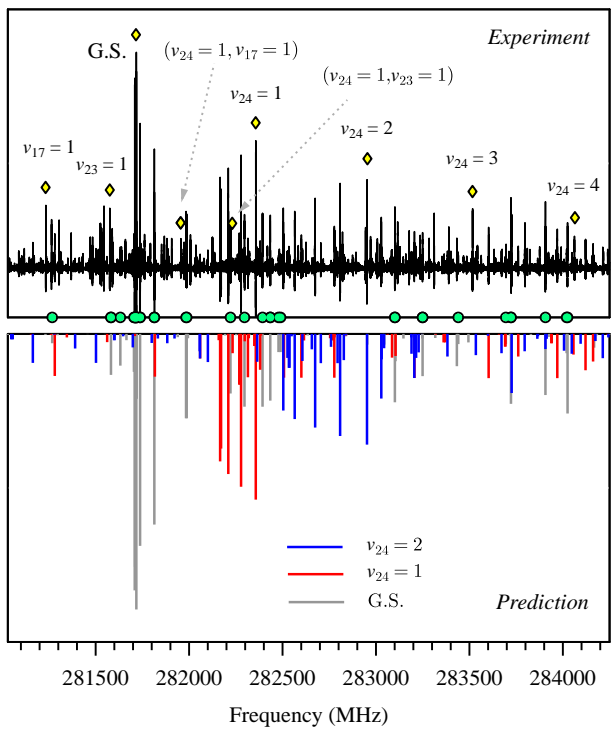}
      \caption{Vibrational satellite pattern of the \textit{syn} conformer in the millimeter wave spectrum of acrylamide. The upper panel shows a pair of degenerate $46_{~1,~46} \leftarrow 45_{~0,~45}$ and $46_{~0,~46} \leftarrow 45_{~1,~45}$ transitions for the ground state (G.S.) and the excited vibrational states (yellow diamonds). The green circle symbols represent the ground state lines included in the fit. The lower panel illustrates a stick reproduction of the ground state and $v_{24}=1,2$ spectra using the spectroscopic constants from Table \ref{T1}.}
      \label{satellites}
   \end{figure}

\section{Rotational spectra and analyses}
{\label{s:analysis}}

\subsection{Syn conformer}

The only available previous microwave study had shown that the \textit{syn} conformer of acrylamide is essentially planar with a large dipole moment component
along the $b$ principal inertial axis ($|\mu_{a}|=$ 0.269(3) D, $|\mu_{b}| =$ 3.42(2) D, $|\mu_{c}| =$ 0.12(24) D; \citealt{Marstokk2000}).
At the initial stage of the analysis, predictions based on the spectroscopic constants from \cite{Marstokk2000} were used.
Strong lines corresponding to pairs of $b$-type $R$-branch transitions $(J+1)_{1,~J+1}\leftarrow J_{0,~J}$ and $(J+1)_{0,~J+1}\leftarrow J_{1,~J}$ were readily assigned as they
are visibly getting closer to each other with increasing $J$ quantum number. They finally
coalesce into very prominent, doubly degenerate lines that stand out in the millimeter wave spectrum in Fig. \ref{features}a.
On the other hand, groups of $b$-type $R$-branch transitions are gradually formed on the way to the sub-millimeter wave region (see Fig. \ref{features}b).
Many $b$-type $Q$-branch transitions could also be identified throughout the spectrum. Finally, $a$-type transitions were searched for but could not be observed. Graphical Loomis–Wood-type plots produced with the AABS package \citep{aabs,Kisiel2012} greatly facilitated the line assignments.
Some of the observed lines were found partially split due to the nuclear quadrupole coupling interactions of a single $^{14}$N nucleus. They were not further taken into consideration due to the distorted line shapes.

We measured more than 2500 lines which encompass the rotational transitions with $J$ and $K_{a}$ quantum numbers up to 78 and 33, respectively. These transitions were ultimately merged with a selection of transitions from \cite{Marstokk2000} which were available in the microwave catalog from Ulm. Like our data set, this selection contained the rotational transitions for which the nuclear quadrupole hyperfine structure was expected to be collapsed. The fits and predictions were made in terms of Watson’s $A$-reduced Hamiltonian in I$^{\text{r}}$-representation \citep{Watson1977}. The broad coverage of transition types and quantum numbers ensured the precise determination of the rotational constants and full sets of the quartic and sextic centrifugal distortion constants.
Their values are collected in Table \ref{T1} while the list of measured transitions is provided in Table \ref{transitions-syn}.

Several excited vibrational states were observed for the \textit{syn} conformer by \cite{Marstokk2000}. The low-frequency C--C torsional mode in this conformer, which under the $C_{s}$ point group is labeled $\nu_{24}$ ($A''$), was found to be a dominating contributor to the vibrational satellite pattern which was also observed in our records (see Fig. \ref{satellites}). Its frequency was estimated to 90(10) cm$^{-1}$ from the microwave relative intensity measurements \citep{Marstokk2000}.
We were able to build upon results from \cite{Marstokk2000} and follow the successive excitation of this mode up to $v_{24}=4$.
However, only $v_{24}=1$ and $v_{24}=2$ could be analyzed using the same Hamiltonian as for the ground state and their spectroscopic constants are given in Table \ref{T1}. The rotational transitions in $v_{24}=3$ and $v_{24}=4$ suffered from perturbations.
These perturbations are not surprising since the manifold of vibrational energy levels become denser above 200 cm$^{-1}$. In the neighborhood of $v_{24}=3$, which under the harmonic approximation can be estimated at 270 cm$^{-1}$, there are first excited states of two other normal vibrational modes: $\nu_{23}$ ($A''$) and $\nu_{17}$ ($A'$) which are associated with NH$_2$ wagging and C=C--C in-plane bending motions, respectively. Their fundamentals were observed at
262.8~cm$^{-1}$  and 307~cm$^{-1}$ using infrared and Raman spectroscopies on gas phase \citep{Kydd1980} and solid phase samples \citep{Duarte2005}, respectively. Microwave relative intensity measurements provided values of 235(40)~cm$^{-1}$ and 307(40)~cm$^{-1}$ for $v_{23}=1$ and $v_{17}=1$, respectively \citep{Marstokk2000}.
Due to the proximity of vibrational energy levels, significant vibration-rotation interactions might be thus manifested in pure rotational spectra of these states. This proximity is apparent from Fig. \ref{satellites} where the same rotational transitions in $v_{24}=3$, $v_{23}=1$, and $v_{17}=1$ reveal very similar relative intensities. Practically the same situation seems to occur also for $v_{24}=4$ which is energetically accompanied by combination states ($v_{24}=1$, $v_{23}=1$) and ($v_{24}=1$, $v_{17}=1$).

We note that we were able to identify the rotational transitions in all these excited vibrational states in our records (see Fig. \ref{satellites}) in agreement with the assignments from \cite{Marstokk2000}. Many lines for these states could be confidently assigned using the Loomis–Wood-type plots but could not be treated in the scope of the semi-rigid rotor Hamiltonian due to clear evidences of perturbations. An inspection of these perturbations suggests a rather complex and time-consuming treatment, probably requiring two three-state Hamiltonian fits. We are planning to discuss this quite intricate undertaking in a separate paper as the outcome will have no bearing on the results reported in the present study.

Finally, we provide in Table \ref{T4} partition functions of \textit{syn} acrylamide needed to estimate the column density. The rotational partition function ($Q_{\text{rot}}$) was evaluated by summation of the Boltzmann factors over the energy levels in the ground vibrational state. We used the SPCAT program \citep{Pickett1991} to undertake this summation numerically employing the spectroscopic constants from Table \ref{T1} and all rotational states up to $J=120$. The vibrational partition function ($Q_{\text{vib}}$) was obtained using the Eq. 3.60 of \cite{Gordy1970} by taking into account the frequencies of twenty-four normal vibrational modes from Table~\ref{vib-modes} of the Appendix.  

\begin{table}
\caption{Partition functions and abundances for the two conformers of acrylamide.}
\label{T4}
\begin{center}
\begin{footnotesize}
\setlength{\tabcolsep}{5.0pt}
\begin{tabular}{ r r r r r r r r}
\hline\hline
   &  \multicolumn{3}{c}{\textit{Syn} conformer} & & \multicolumn{3}{c}{\textit{Skew} conformer} \\
\cline{2-4}
\cline{6-8}
$T$ (K) &  $Q_{\text{rot}}$ & $Q_{\text{vib}}$ & (\%)\tablefootmark{a} & & $Q_{\text{rot}}$\tablefootmark{b} & $Q_{\text{vib}}$ & (\%) \\
\hline
   300.000    &  74866.65  &  8.11 &  89 &  &  10980.80   &  6.65 &  11  \\
   225.000    &  48631.40  &  3.97 &  95 &  &   2962.15   &  3.37 &   5  \\
   150.000    &  26465.75  &  2.08 &  99 &  &    278.14   &  1.86 &   1  \\
    75.000    &   9356.70  &  1.23 & 100 &  &      0.51   &  1.18 &   0  \\
    37.500    &   3309.64  &  1.03 & 100 &  &      0.00   &  1.02 &   0  \\
    18.750    &   1171.54  &  1.00 & 100 &  &      0.00   &  1.00 &   0  \\
     9.375    &    415.26  &  1.00 & 100 &  &      0.00   &  1.00 &   0  \\
\hline
\end{tabular}
\end{footnotesize}
\end{center}
\tablefoot{
\tablefoottext{a}{Conformer abundance calculated as $Q_{\text{tot}}^{\mathit{syn}}/(Q_{\text{tot}}^{\mathit{syn}}+Q_{\text{tot}}^{\mathit{skew}})$ where $Q_{\text{tot}}$ is the total partition function for a given conformer obtained as the product of $Q_{\text{rot}}$ and $Q_{\text{vib}}$.}
\tablefoottext{b}{Accounts for the ground state tunneling doublet corrected for the energy of the \textit{skew} conformer of $E_{\mathit{skew}} =$ 543~cm$^{-1}$ taken from \cite{Marstokk2000}.}}
\end{table}

\begin{table*}
\caption{Spectroscopic constants for \textit{skew} acrylamide in the ground state tunneling doublet ($A$-reduction, I$^{\text{r}}$-representation).}
\label{T3}
\begin{center}
\begin{footnotesize}
\setlength{\tabcolsep}{8.5pt}
\begin{tabular}{ l r r r r r}
\hline\hline
 &    \multicolumn{2}{c}{This work} &  &  \multicolumn{2}{c}{\cite{Marstokk2000}}\tablefootmark{a} \\
\cline{2-3}
\cline{5-6}
 &    $0^+$ &  $0^-$ &  &  $0^+$ &  $0^-$\\
\hline
$A                   $  /               MHz     & 10049.46072 (34)\tablefootmark{b}     &   9996.61018 (39)     & & 10049.549 (22)  &  10002.8 (5)    \\
$B                   $  /               MHz     &  4287.87861 (14)                      &   4291.77177 (19)     & &  4287.924 (12)  &   4292.160 (24) \\
$C                   $  /               MHz     &  3035.794577 (67)                     &   3050.747935 (98)    & &  3035.806 (12)  &   3050.385 (33) \\
$\Delta_{J}          $  /               kHz     &     0.990767 (85)                     &      0.95662 (11)     & &  1.14 (9)       &   1.15 (16)     \\
$\Delta_{JK}         $  /               kHz     &     3.47106 (38)                      &      4.24560 (48)     & &  4.20 (12)      &   3.35 (18)     \\
$\Delta_{K}          $  /               kHz     &     7.0201 (13)                       &      5.0704 (15)      & &  8.0 (6)        &   8.0           \\
$\delta_{J}          $  /               kHz     &     0.166200 (44)                     &      0.194935 (60)    & &  0.283 (9)      &   0.28          \\
$\delta_{K}          $  /               kHz     &     5.03612 (65)                      &      4.64196 (99)     & &  3.23 (15)      &   3.2           \\

$\Phi_{J}            $  /                Hz     &     0.002027 (27)                     &      0.000986 (39)    & &   ...      &  ...     \\
$\Phi_{JK}           $  /                Hz     &   --0.00838 (29)                      &    --0.00530 (29)     & &   ...      &  ...     \\
$\Phi_{KJ}           $  /                Hz     &   --0.2776 (11)                       &    --0.1182 (12)      & &   ...      &  ...     \\
$\Phi_{K}            $  /                Hz     &     0.1495 (16)                       &      0.0781 (18)      & &   ...      &  ...     \\
$\phi_{J}            $  /                Hz     &   --0.000622 (14)                     &    --0.000261 (19)    & &   ...      &  ...     \\
$\phi_{JK}           $  /                Hz     &     0.02621 (24)                      &      0.00500 (34)     & &   ...      &  ...     \\
$\phi_{K}            $  /                Hz     &   --0.1939 (23)                       &    --0.0737 (26)      & &   ...      &  ...     \\

$\Delta E            $  /               MHz     &     \multicolumn{2}{c}{415050.244 (55)}                       & &   \multicolumn{2}{c}{...}     \\
$\Delta E            $  /         cm$^{-1}$     &     \multicolumn{2}{c}{13.8445859 (18)}                       & &   \multicolumn{2}{c}{...}     \\
$F_{bc}              $  /               MHz     &     \multicolumn{2}{c}{4.08214 (95)}                          & &   \multicolumn{2}{c}{...}     \\
$F_{bc}^J            $  /               kHz     &     \multicolumn{2}{c}{--0.14465 (66)}                        & &   \multicolumn{2}{c}{...}     \\
$F_{bc}^K            $  /               kHz     &     \multicolumn{2}{c}{ 1.1873 (37)}                          & &   \multicolumn{2}{c}{...}     \\
$F_{bc}^{JJ}         $  /                Hz     &     \multicolumn{2}{c}{0.001568 (53)}                         & &   \multicolumn{2}{c}{...}     \\
$F_{bc}^{JK}         $  /                Hz     &     \multicolumn{2}{c}{0.0624 (25)}                           & &   \multicolumn{2}{c}{...}     \\
$F_{ac}              $  /               MHz     &     \multicolumn{2}{c}{--1.2209 (94)}                         & &   \multicolumn{2}{c}{...}     \\
$F_{ac}^J            $  /               kHz     &     \multicolumn{2}{c}{--0.2239 (41)}                         & &   \multicolumn{2}{c}{...}     \\
$F_{ac}^K            $  /               kHz     &     \multicolumn{2}{c}{3.407 (25)}                            & &   \multicolumn{2}{c}{...}     \\
$J_{\text{min}}/J_{\text{max}}$                 &    4 / 76         &   6 / 77                                  & &  4 / 27 & 6 / 8 \\
$K_{a}^{\text{min}}/K_{a}^{\text{max}}$         &   0 / 24          &   0 / 24                                  & &  0 / 8  & 2 / 7 \\
$N$\tablefootmark{c}                            &   1419            &   1255                                    & &  53     & 14    \\
$\sigma_{\text{fit}}$\tablefootmark{d}/ MHz     &   0.040           &   0.035                                   & &  0.196  & 0.129 \\
$\sigma_{\text{w}}$\tablefootmark{e}            &   1.03            &   1.08                                    & &  ...    & ...   \\
\hline
\end{tabular}
\end{footnotesize}
\end{center}
\tablefoot{
\tablefoottext{a}{The ground state and the excited C--C torsional state in \cite{Marstokk2000} correspond to $0^+$ and $0^-$ states, respectively.}
\tablefoottext{b}{The numbers in parentheses are the parameter uncertainties in units of the last decimal digit. Their values are close to 1$\sigma$ standard uncertainties (67\% confidence level) because the unitless (weighted) deviation of the fit is close to 1.0. SPFIT/SPCAT program package \citep{Pickett1991} was used for the analysis.}
\tablefoottext{c}{Number of distinct frequency lines in the fit.}
\tablefoottext{d}{Root mean square deviation of the fit.}
\tablefoottext{e}{Unitless (weighted) deviation of the fit.}}
\end{table*}

\subsection{Skew conformer}
\label{skew}

Spectroscopic constants from \cite{Marstokk2000} were used to search for rotational transitions of the less stable \textit{skew} conformer in our records.
Since the candidate lines were significantly weaker and were displaced by a few MHz from the predicted positions, a double check for the correct assignment in the very dense spectrum
was the simultaneous observation of $a$-type and $b$-type transitions. Unlike the \textit{syn} form, which possesses a sizable dipole moment component only along the $b$ principal axis,
the \textit{skew} conformer is endowed with large dipole moment components along the $a$ and $b$ axes ($|\mu_{a}|=$ 2.81 D, $|\mu_{b}| =$ 3.02 D, $|\mu_{c}| =$ 1.00 D as calculated by \citealt{Marstokk2000} at the MP2=full/cc-pVTZ level of theory). For certain values of $J$ this leads to the observations of easily discernible quartets in which the two $b$-type transitions straddle the pair of $a$-type transitions.
This is clearly illustrated in Fig. \ref{features}c which shows a small part of the spectrum at 154 GHz.

      \begin{figure}[ht]
   \centering
   \includegraphics[trim = 1mm 1mm 1mm 1mm, clip, width=7.8cm]{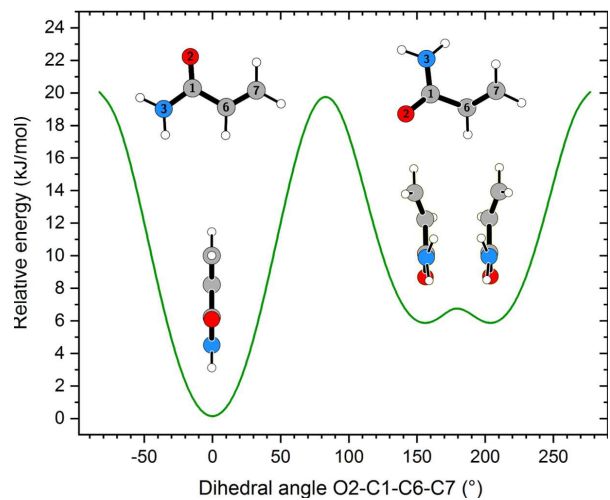}
      \caption{Potential energy surface scan for the skeletal torsion calculated at the B3LYP/cc-pVTZ level of theory using the scan option of Gaussian 16 \citep{g16}. The energies were calculated in steps of 10$^{\circ}$ of the dihedral angle O2--C1--C6--C7 while the other structural parameters were optimized under the \textit{verytight} convergence criterion. The global minimum belongs to the planar \textit{syn} conformer while for the higher-energy conformer two minima can be identified on the surface which correspond to two equivalent \textit{skew} frameworks connected with a barrier of around 1 kJ mol$^{-1}$ (84 cm$^{-1}$).
      }
      \label{doublemin}
   \end{figure}

In this study, we assigned transitions belonging to the ground state and a higher-frequency satellite matching with what was assumed to be the first
excited C–C torsional state in \cite{Marstokk2000}. Relative intensity measurements undertaken by \cite{Marstokk2000} placed this state into the energy window between 71 and 111 cm$^{-1}$ above the ground state. On the other hand, our records indicate that this state lies significantly lower; its rotational lines apparently presented intensities very similar to those of the corresponding lines in the ground state. An examination of the Loomis–Wood-type plots made it further possible to identify several level crossing perturbations. They were observable in mirror-image form for the ground state and the satellite lines, presenting evidence
that these two states are in close proximity and are in mutual interaction.

An explanation for the observation of such an interacting pair of states could be a double-minimum potential for \textit{skew} acrylamide which we subsequently confirmed by means of a potential energy surface scan for the rotation about the central C--C bond. As shown in Fig. \ref{doublemin}, two minima corresponding to two symmetrically equivalent non-planar \textit{skew} structures are separated by a small barrier. In this scenario, the ground vibrational state, normally labeled $v = 0$, is expected to be observed in the form of a doublet, the members of which are designated $0^+$ and $0^-$. These two states are often affected by vibration-rotation interactions as shown, for example, for glycinamide \citep{Kisiel2021}, cyanamide \citep{Kisiel2013}, 2-hydroxyacetonitrile \citep{Margules2017}, and $aGg'$ ethylene glycol \citep{Christen2003}.
The identification of characteristic mirror-image perturbation spikes for various $K_{a}$, i.e. frequency shifts with respect to the unperturbed positions (see, e.g., \citealt{Kisiel2013}), allowed us to unambiguously assign the higher-frequency satellite to the higher-energy $0^-$ state. Furthermore, their precise location in $J$ and $K_a$ allowed rather confident estimation of the energy difference $\Delta E = E^{(0^-)}-E^{(0^+)} \approx$  13.8 cm$^{-1}$. This value was used as an estimate in the fitting procedure that employed a two-state Hamiltonian in the following matrix form
\begin{equation}
H_{\mathrm{eff}} = \begin{pmatrix}
    H_{\mathrm{rot}}^{(0^+)} & H_{\mathrm{cor}}^{(0^{+},0^{-})}  \\
     H_{\mathrm{cor}}^{(0^{+},0^{-})} & H_{\mathrm{rot}}^{(0^+)} + \Delta E
    \end{pmatrix}
=
\begin{pmatrix}
    H_{\mathrm{rot}}^{(0^+)} & H_{\mathrm{cor}}^{(0^{+},0^{-})}  \\
     H_{\mathrm{cor}}^{(0^{+},0^{-})} & H_{\mathrm{rot}}^{(0^-)}
    \end{pmatrix}
\end{equation}
where $H_{\mathrm{rot}}^{(0^+)}$ and $H_{\mathrm{rot}}^{(0^-)}$ represent Watson's $A$-reduced semi-rigid rotor Hamiltonians
in I$^{\mathrm{r}}$-representations \citep{Watson1977} for $0^{+}$ and $0^{-}$ states, respectively.
The off-diagonal coupling term $H_{\mathrm{cor}}^{(0^{+},0^{-})}$ is based on a reduced axis system (RAS) approach of \cite{Pickett1972} which was used to treat perturbations in molecular systems with symmetric double-minimum potentials such as cyanamide \citep{Kisiel2013}, propanal \citep{Zingsheim2017}, and methoxymethanol \citep{Motiyenko2018} and takes the form
\begin{eqnarray}
 H_{\rm cor}^{(0^+,0^-)}  & = &  (F_{bc}+ F_{bc}^J J^2 + F_{bc}^{K} J_z^2 +\dots) [J_b, J_c]_+ \nonumber\\
                          &   &  (F_{ac}+ F_{ac}^J J^2 + F_{ac}^{K} J_z^2 +\dots) [J_a, J_c]_+
\label{cor}
\end{eqnarray}
where $F_{bc}$ and $F_{ac}$ are the interaction parameters. Their centrifugal distortion expansion terms ($F_{bc}^J$, $F_{bc}^K$, ...) were systematically explored during the fitting procedure. A rather satisfactory fit, which is based on more than 2600 lines (our lines and those from \citealt{Marstokk2000}), was eventually reached with the choice for the parameter set given in Table \ref{T3}. The list of measured rotational transitions is provided in Table \ref{transitions-skew} of the Appendix.
Since the perturbations affect not only line frequencies but also line intensities \citep{Christen2003,Kisiel2021}, we tested the relative signs for the $F$ parameters with respect to those of the dipole moment components. The sign combination in Table \ref{T3} with positive $\mu_{a}$ and $\mu_{b}$ values reproduce the intensity alternations of perturbed transitions in the spectrum. 
We note that the same result is obtained if the signs of all $F$ parameters are reversed or those of all dipole moment components.

Final remarks concern the quartic and sextic centrifugal distortion constants in Table \ref{T3}.
Their values in $0^+$ and $0^-$ tunneling states are generally close to each other, however, a difference might be perceptible, for example, for $\Delta_{K}$.
The main difficulty here was the high spectral density caused by the existence of the more stable \textit{syn} conformer and its rotational lines in many excited vibrational states. It was sometimes rather difficult to evaluate whether the lines are obscured by these features or are still perturbed or simply are not observed due to their weakness and our limited  signal to noise ratio. It is therefore possible that the perturbation contributions are treated incompletely. Nevertheless, the measured data set for \textit{skew} acrylamide is relatively large and is reproduced near the experimental uncertainty using the spectroscopic parameters from Table \ref{T3}.

\begin{table*}[!ht]
 \begin{center}
 \caption{
 Parameters of our best-fit LTE model of acetamide toward Sgr~B2(N1S), and upper limits for propiolamide, acrylamide, and propionamide.
}
 \label{t:coldens_n1s}
 \vspace*{-1.2ex}
 \begin{tabular}{llcrccccccr}
 \hline\hline
 \multicolumn{2}{c}{Molecule} & \multicolumn{1}{c}{Status\tablefootmark{a}} & \multicolumn{1}{c}{$N_{\rm det}$\tablefootmark{b}} & \multicolumn{1}{c}{$\theta_{\rm s}$\tablefootmark{c}} & \multicolumn{1}{c}{$T_{\mathrm{rot}}$\tablefootmark{d}} & \multicolumn{1}{c}{$N$\tablefootmark{e}} & \multicolumn{1}{c}{$F_{\rm vib}$\tablefootmark{f}} & \multicolumn{1}{c}{$\Delta V$\tablefootmark{g}} & \multicolumn{1}{c}{$V_{\mathrm{off}}$\tablefootmark{h}} & \multicolumn{1}{c}{$\frac{N_{\rm ref}}{N}$\tablefootmark{i}} \\
  & & & & \multicolumn{1}{c}{\small ($''$)} & \multicolumn{1}{c}{\small (K)} & \multicolumn{1}{c}{\small (cm$^{-2}$)} & & \multicolumn{1}{c}{\small (km~s$^{-1}$)} & \multicolumn{1}{c}{\small (km~s$^{-1}$)} & \\
 \hline
 \textit{Acetamide} & CH$_3$C(O)NH$_2$\tablefootmark{(j)}$^\star$ & d & 153 &  2.0 &  160 &  4.1 (17) & 1.16 & 5.0 & $0.0$ &       1 \\
\hline
 \textit{Propiolamide} & HCCC(O)NH$_2$\tablefootmark{(k)} & n & 0 &  2.0 &  160 & $<$  8.5 (15) & 1.18 & 5.0 & $0.0$ & $>$      48 \\
\hline
 \textit{Acrylamide} & \textit{syn}-C$_2$H$_3$C(O)NH$_2$ & n & 0 &  2.0 &  160 & $<$  1.6 (16) & 2.26 & 5.0 & $0.0$ & $>$      26 \\
\hline
 \textit{Propionamide} & C$_2$H$_5$C(O)NH$_2$ & n & 0 &  2.0 &  160 & $<$  2.9 (16) & 4.16 & 5.0 & $0.0$ & $>$      14 \\
\hline
 \end{tabular}
 \end{center}
 \vspace*{-2.5ex}
 \tablefoot{
 \tablefoottext{a}{d: detection, n: nondetection.}
 \tablefoottext{b}{Number of detected lines \citep[conservative estimate, see Sect.~3 of][]{Belloche16}. One line of a given species may mean a group of transitions of that species that are blended together.}
 \tablefoottext{c}{Source diameter (FWHM).}
 \tablefoottext{d}{Rotational temperature.}
 \tablefoottext{e}{Total column density of the molecule. $x$ ($y$) means $x \times 10^y$.}
 \tablefoottext{f}{Correction factor that was applied to the column density to account for the contribution of vibrationally excited states, in the cases where this contribution was not included in the partition function of the spectroscopic predictions.}
 \tablefoottext{g}{Linewidth (FWHM).}
 \tablefoottext{h}{Velocity offset with respect to the assumed systemic velocity of Sgr~B2(N1S), $V_{\mathrm{sys}} = 62$ km~s$^{-1}$.}
 \tablefoottext{i}{Column density ratio, with $N_{\rm ref}$ the column density of the previous reference species marked with a $\star$.}
 \tablefoottext{j}{The parameters were derived from the ReMoCA survey by \citet{Belloche2019}.}
 \tablefoottext{k}{The upper limit was derived from the ReMoCA survey by \citet{Alonso2021}.}
 }
 \end{table*}

\begin{table*}[!ht]
 \begin{center}
 \caption{
 Parameters of our best-fit LTE model of acetamide toward Sgr~B2(N2), and upper limits for propiolamide, acrylamide, and propionamide.
}
 \label{t:coldens_n2}
 \vspace*{-1.2ex}
 \begin{tabular}{llcrccccccr}
 \hline\hline
 \multicolumn{2}{c}{Molecule} & \multicolumn{1}{c}{Status\tablefootmark{a}} & \multicolumn{1}{c}{$N_{\rm det}$\tablefootmark{b}} & \multicolumn{1}{c}{$\theta_{\rm s}$\tablefootmark{c}} & \multicolumn{1}{c}{$T_{\mathrm{rot}}$\tablefootmark{d}} & \multicolumn{1}{c}{$N$\tablefootmark{e}} & \multicolumn{1}{c}{$F_{\rm vib}$\tablefootmark{f}} & \multicolumn{1}{c}{$\Delta V$\tablefootmark{g}} & \multicolumn{1}{c}{$V_{\mathrm{off}}$\tablefootmark{h}} & \multicolumn{1}{c}{$\frac{N_{\rm ref}}{N}$\tablefootmark{i}} \\
  & & & & \multicolumn{1}{c}{\small ($''$)} & \multicolumn{1}{c}{\small (K)} & \multicolumn{1}{c}{\small (cm$^{-2}$)} & & \multicolumn{1}{c}{\small (km~s$^{-1}$)} & \multicolumn{1}{c}{\small (km~s$^{-1}$)} & \\
 \hline
 \textit{Acetamide} & CH$_3$C(O)NH$_2$\tablefootmark{(j)}$^\star$ & d & 23 &  0.9 &  180 &  1.4 (17) & 1.23 & 5.0 & $1.5$ &       1 \\
\hline
 \textit{Propiolamide} & HCCC(O)NH$_2$\tablefootmark{(k)} & n & 0 &  0.9 &  180 & $<$  1.0 (16) & 1.30 & 5.0 & $0.0$ & $>$      13 \\
\hline
 \textit{Acrylamide} & \textit{syn}-C$_2$H$_3$C(O)NH$_2$ & n & 0 &  0.9 &  180 & $<$  2.1 (16) & 2.67 & 5.0 & $0.0$ & $>$     6.3 \\
\hline
 \textit{Propionamide} & C$_2$H$_5$C(O)NH$_2$ & n & 0 &  0.9 &  180 & $<$  4.3 (16) & 5.06 & 5.0 & $0.0$ & $>$     3.1 \\
\hline
 \end{tabular}
 \end{center}
 \vspace*{-2.5ex}
 \tablefoot{
 \tablefoottext{a}{d: detection, n: nondetection.}
 \tablefoottext{b}{Number of detected lines \citep[conservative estimate, see Sect.~3 of][]{Belloche16}. One line of a given species may mean a group of transitions of that species that are blended together.}
 \tablefoottext{c}{Source diameter (FWHM).}
 \tablefoottext{d}{Rotational temperature.}
 \tablefoottext{e}{Total column density of the molecule. $x$ ($y$) means $x \times 10^y$.}
 \tablefoottext{f}{Correction factor that was applied to the column density to account for the contribution of vibrationally excited states, in the cases where this contribution was not included in the partition function of the spectroscopic predictions.}
 \tablefoottext{g}{Linewidth (FWHM).}
 \tablefoottext{h}{Velocity offset with respect to the assumed systemic velocity of Sgr~B2(N2), $V_{\mathrm{sys}} = 74$ km~s$^{-1}$.}
 \tablefoottext{i}{Column density ratio, with $N_{\rm ref}$ the column density of the previous reference species marked with a $\star$.}
 \tablefoottext{j}{The parameters were derived from the EMoCA survey by \citet{Belloche2017}.}
 \tablefoottext{k}{The upper limit was derived from the ReMoCA survey by \citet{Alonso2021}.}
 }
 \end{table*}

\section{Search for acrylamide and related molecules toward Sgr~B2(N)}
\label{s:astro}

\subsection{Observations}
\label{ss:observations}

We used the imaging spectral line survey ReMoCA (Reexploring Molecular
Complexity with ALMA) performed with the Atacama Large
Millimeter/submillimeter Array (ALMA) to search
for acrylamide toward Sgr~B2(N). The observational strategy and the method
employed to reduce the data of this survey were described in
\citet{Belloche2019}. We summarize here the main features. The survey extends
from 84.1~GHz to 114.4~GHz with a spectral resolution of 488~kHz (1.7 to
1.3~km~s$^{-1}$). It achieved a median angular resolution (HPBW) of
0.6$\arcsec$, with values varying between $\sim$0.3$\arcsec$ and
$\sim$0.8$\arcsec$. The median resolution corresponds to $\sim$4900~au at the
distance of Sgr~B2 \citep[8.2~kpc,][]{Reid19}. The interferometric
observations were centered on the equatorial position
($\alpha, \delta$)$_{\rm J2000}$=
($17^{\rm h}47^{\rm m}19{\fs}87, -28^\circ22'16{\farcs}0$) that is located half-way
between Sgr~B2(N1) and Sgr~B2(N2), the two main hot molecular cores of
Sgr~B2(N). These hot cores are separated by 4.9$\arcsec$ or
$\sim$0.2~pc in projection onto the plane of the sky. The survey achieved a
median sensitivity per spectral channel of 0.8~mJy~beam$^{-1}$ (rms), with
values ranging between 0.35~mJy~beam$^{-1}$ and 1.1~mJy~beam$^{-1}$.

Our search for acrylamide in Sgr~B2(N) followed the same strategy as our
search for propiolamide, HCCC(O)NH$_2$, reported in \citet{Alonso2021}. We
focused the search toward the following two positions: the offset position
Sgr~B2(N1S) located at ($\alpha, \delta$)$_{\rm J2000}$=
($17^{\rm h}47^{\rm m}19{\fs}870$, $-28^\circ22\arcmin19{\farcs}48$) and the
secondary hot core Sgr~B2(N2) at ($\alpha, \delta$)$_{\rm J2000}$=
($17^{\rm h}47^{\rm m}19{\fs}863$, $-28^\circ22\arcmin13{\farcs}27$). The former
position was chosen by \citet{Belloche2019}. With its location about 1$\arcsec$
to the South of the main hot core Sgr~B2(N1), its continuum emission has a
lower opacity, which allows us to look deeper into the molecular content of
Sgr~B2(N1). Compared to \citet{Belloche2019}, we employed a more recent version
of our data set for which we have improved the splitting of the continuum and
line emission as reported in \citet{Melosso20}.

We modeled the spectra of Sgr~B2(N1S) and Sgr~B2(N2) with the software Weeds
\citep[][]{Maret11} under the assumption of local thermodynamic equilibrium
(LTE), which is appropriate given the high densities that characterize
Sgr~B2(N)'s hot cores \citep[$>1 \times 10^{7}$~cm$^{-3}$, see][]{Bonfand19}. A
best-fit synthetic spectrum of each molecule was derived separately, and then
the contributions of all identified molecules were added together. Each species
was modeled with a set of five parameters: size of the emitting region
($\theta_{\rm s}$), column density ($N$), temperature ($T_{\rm rot}$), linewidth
($\Delta V$), and velocity offset ($V_{\rm off}$) with respect to the assumed
systemic velocity of the source, $V_{\rm sys}=62$~km~s$^{-1}$ for Sgr~B2(N1S) and
$V_{\rm sys}= 74$~km~s$^{-1}$ for Sgr~B2(N2).

\begin{table*}[!ht]
 \begin{center}
 \caption{
 Column density upper limit derived for propionamide toward the offset position (1.58$\arcsec$,$-$2.72$\arcsec$) with respect to the phase center.
}
 \label{t:coldens_p158m272}
 \vspace*{-1.2ex}
 \begin{tabular}{lcrcccccc}
 \hline\hline
 \multicolumn{1}{c}{Molecule} & \multicolumn{1}{c}{Status\tablefootmark{a}} & \multicolumn{1}{c}{$N_{\rm det}$\tablefootmark{b}} & \multicolumn{1}{c}{$\theta_{\rm s}$\tablefootmark{c}} & \multicolumn{1}{c}{$T_{\mathrm{rot}}$\tablefootmark{d}} & \multicolumn{1}{c}{$N$\tablefootmark{e}} & \multicolumn{1}{c}{$F_{\rm vib}$\tablefootmark{f}} & \multicolumn{1}{c}{$\Delta V$\tablefootmark{g}} & \multicolumn{1}{c}{$V_{\mathrm{off}}$\tablefootmark{h}} \\ 
  & & & \multicolumn{1}{c}{\small ($''$)} & \multicolumn{1}{c}{\small (K)} & \multicolumn{1}{c}{\small (cm$^{-2}$)} & & \multicolumn{1}{c}{\small (km~s$^{-1}$)} & \multicolumn{1}{c}{\small (km~s$^{-1}$)} \\ 
 \hline
 C$_2$H$_5$C(O)NH$_2$ & n & 0 &  2.3 &  150 & $<$  1.3 (16) & 3.77 & 4.2 & $0.0$ \\ 
\hline 
 \end{tabular}
 \end{center}
 \vspace*{-2.5ex}
 \tablefoot{
 \tablefoottext{a}{n: nondetection.}
 \tablefoottext{b}{Number of detected lines.}
 \tablefoottext{c}{Source diameter (FWHM).}
 \tablefoottext{d}{Rotational temperature.}
 \tablefoottext{e}{Total column density of the molecule. $x$ ($y$) means $x \times 10^y$.}
 \tablefoottext{f}{Correction factor that was applied to the column density to account for the contribution of vibrationally excited states, in the cases where this contribution was not included in the partition function of the spectroscopic predictions.}
 \tablefoottext{g}{Linewidth (FWHM).}
 \tablefoottext{h}{Velocity offset with respect to the assumed systemic velocity, $V_{\mathrm{sys}} = 63$ km~s$^{-1}$.}
 }
 \end{table*}

\subsection{Nondetection of acrylamide}
\label{ss:acrylamide_nondetection}

To guide our search for the \textit{syn} conformer of acrylamide in the ReMoCA spectra
of Sgr~B2(N1S) and Sgr~B2(N2), we computed LTE synthetic spectra of this
molecule on the basis of the parameters derived for acetamide,
CH$_3$C(O)NH$_2$, by \citet{Belloche2019} and \citet{Belloche2017}, respectively.
Only the column density of acrylamide was kept as a free parameter. We
searched for rotational lines in its vibrational ground state, $\varv=0$, and
in its vibrationally excited states $\varv_{24}=1$ and $\varv_{24}=2$. No
evidence for emission of acrylamide was found toward either source. The
nondetection toward Sgr~B2(N1S) and Sgr~B2(N2) is illustrated in
Figs.~\ref{f:spec_c2h3conh2-s_ve0_n1s}--\ref{f:spec_c2h3conh2-s_v24e1_n1s} and
Figs.~\ref{f:spec_c2h3conh2-s_ve0_n2}--\ref{f:spec_c2h3conh2-s_v24e1_n2},
respectively, and the upper limits to the column density of acrylamide are
indicated in Tables~\ref{t:coldens_n1s} and \ref{t:coldens_n2}, respectively.
The tables also recall the parameters that we previously obtained for acetamide
and propiolamide. We did not take into account
the contribution of the \textit{skew} conformer of acrylamide to its partition
function. This conformer is observed as a tunneling doublet about 540~cm$^{-1}$ above the \textit{syn} form (see Sect.\ref{skew}).
Relative abundances of each conformer in Table \ref{T4} show that it contributes less than 3\% to the partition function at 160--180~K, which does not have a
significant impact on the upper limits reported for acrylamide in
Tables~\ref{t:coldens_n1s} and \ref{t:coldens_n2}.

\subsection{Nondetection of propionamide}
\label{ss:propionamide_nondetection}

We also searched for propionamide, C$_2$H$_5$C(O)NH$_2$, which has a fully
saturated alkyl group and is structurally related to acrylamide. We used the
spectroscopic predictions for its vibrational ground state, $\varv=0$,
and its first vibrationally excited state, $\varv_{30}=1$, published in
\citet{Li2021}. We did not find any evidence for this molecule toward either
Sgr~B2(N1S) or Sgr~B2(N2), as illustrated in
Figs.~\ref{f:spec_c2h5conh2_ve0_n1s}--\ref{f:spec_c2h5conh2_v30e1_n1s} and
\ref{f:spec_c2h5conh2_ve0_n2}--\ref{f:spec_c2h5conh2_v30e1_n2},
respectively. Tables~\ref{t:coldens_n1s} and \ref{t:coldens_n2}, respectively,
report the upper limits that we derived for its column density.

\citet{Li2021} recently claimed the tentative detection of propionamide on the
basis of the ReMoCA data set, which they downloaded from the ALMA archive and
reduced themselves, toward a position located to the East of Sgr B2(N1)
at ($\alpha, \delta$)$_{\rm J2000}$=
($17^{\rm h}47^{\rm m}19{\fs}99, -28^\circ22'18{\farcs}72$). This position corresponds to
an equatorial coordinate offset of (1.58$\arcsec$, $-$2.72$\arcsec$) with respect to the
phase center. We also searched for propionamide toward this position using our
own reduced version of the ReMoCA survey. We computed LTE synthetic spectra
assuming the same source size, temperature, linewidth, and velocity as
\citet{Li2021} and kept only the column density as a free parameter.
We did not find any evidence for propionamide toward this position either. The
nondetection toward this position is illustrated in
Figs.~\ref{f:spec_c2h5conh2_ve0_p158m272} and
\ref{f:spec_c2h5conh2_v30e1_p158m272} and the upper limit to its column
density is reported in Table~\ref{t:coldens_p158m272}.

\section{Discussion}
\label{s:discussion}

\subsection{Laboratory spectroscopy of acrylamide}

The spectroscopic results from Sect. \ref{s:analysis} represent a substantial improvement in the laboratory characterization
of the rotational spectrum of acrylamide. We have significantly extended the frequency coverage
with over 8500 new transitions assigned and analyzed. The derived spectroscopic constants reproduce the assigned transitions within the experimental uncertainties on average and allow to confidently predict the rotational transitions of acrylamide over a broad range of frequencies. Of particular interest in this work was the observational window from 84.1 to 114.4 GHz in which the rotational transitions already revealed significant differences between the experimental frequencies and spectral predictions based on the original microwave work of \cite{Marstokk2000}. For example, ground state $K_a = 1 \leftarrow 0$ transitions of the \textit{syn} conformer were shifted almost by 2 MHz at the predicted uncertainty around 100 kHz.
Since the \textit{syn} conformer represents the global minimum on the potential energy landscape of acrylamide, it is the most relevant
species for observational purposes. Its ground state catalog derived from this work represents highly accurate observational reference and will be available in the Cologne Database for Molecular Spectroscopy (CDMS) to support future searches for acrylamide in space, e.g. in warm or lukewarm environments.

A comparison between the derived spectroscopic parameters of the \textit{syn} conformer and those from the previous microwave study is shown in Table \ref{T1}.
The most immediate conclusion to be extracted from this table is that the accuracy of these parameters is significantly improved for both the ground state and the two low-lying excited vibrational states. However, it is noticeable that while the rotational and quartic centrifugal distortion constants agree with those from \cite{Marstokk2000}, significant discrepancies are found for the sextic constants. These discrepancies are very likely attributed to incorrectly treated nuclear quadrupole hyperfine splitting, and to a lesser extent to a few misassignments in the earlier study. It has been revealed that many transitions need to be assigned to
frequencies of the overlapping $F = J \pm 1$ hyperfine components,
instead of the hyperfine-free frequencies. Rather large differences in observed minus calculated frequencies for these transitions were then artificially fit to sextic centrifugal distortion constants. However, this is no longer possible when a significantly larger data set is combined with these transitions; they clearly appear as outliers in the fit. Taking the \textit{syn} conformer ground state as an example, Table \ref{Tcomparison} shows that when the incriminated lines are removed from the data set of \cite{Marstokk2000}, the remaining transitions can be well fitted using two sextic constants. Their values are close to their counterparts determined solely from the data set measured in this work. Table \ref{Tcomparison}. also shows that the microwave data below 60 GHz have no impact on the results of the global fit presented here. It can be explained by the fact that the access of quantum numbers of the present and previous data sets are similar. Nevertheless, the low-frequency data are maintained in the global fit for completeness.

An interesting phenomenon in this study is tunneling splitting for the \textit{skew} conformer that has been observed for this species for the first time. The results of the fit for the ground state tunneling doublet are provided in Table \ref{T3} alongside the results of the previous work. In that work, the two states were analyzed separately and were assigned to the ground state (here $0^+$) and the excited C--C torsional state (here $0^-$). The spectroscopic constants are significantly improved in terms of accuracy and are consistent with those reported previously. Differences are attributable to substantially larger data set treated in this work, containing a variety of transitions. For example, only 14 \textit{a}-type $R$-branch transitions were measured for the $0^-$ state by \cite{Marstokk2000} while over 1200 lines encompassing \textit{a}-type and \textit{b}-type $R$-branch and \textit{b}-type $Q$-branch transitions were measured here. Although this conformer is less important for observational purposes, we made a considerable progress in the characterization of its rotational spectrum and deliver first experimental information on tunneling splitting for this species.

\subsection{Comparison to other molecules in Sgr~B2(N)}
\label{ss:discussion_acrylamide}

The nondetection of acrylamide reported in
Sect.~\ref{ss:acrylamide_nondetection} implies that acrylamide is at least
$\sim$26 and $\sim$6 times less abundant than acetamide toward Sgr~B2(N1S)
and Sgr~B2(N2), respectively.
For comparison, vinyl cyanide, C$_2$H$_3$CN, is about 5 times less abundant
than methyl cyanide, CH$_3$CN, toward Sgr~B2(N2) \citep[][]{Belloche16} and a
preliminary analysis of the ReMoCA survey yields a similar (slightly smaller)
difference for Sgr~B2(N1S). Both pairs of molecules,
the amides C$_2$H$_3$C(O)NH$_2$/CH$_3$C(O)NH$_2$ and the cyanides
C$_2$H$_3$CN/CH$_3$CN, share the same structural difference (an unsaturated
C$_2$H$_3$ group replaced with a saturated CH$_3$ group). The upper limit to
the abundance ratio C$_2$H$_3$C(O)NH$_2$/CH$_3$C(O)NH$_2$ toward Sgr~B2(N2) is
not constraining compared to the cyanide ratio, but toward Sgr~B2(N1S), the
amide ratio is at least 5 times smaller than the cyanide ratio.

\subsection{Nondetection of propionamide}
\label{ss:discussion_propionamide}

Some of the ALMA spectra shown in Figs.~\ref{f:spec_c2h5conh2_ve0_p158m272} and
\ref{f:spec_c2h5conh2_v30e1_p158m272} differ from the spectra shown in
Figs.~5--7 of \citet{Li2021}. At least two reasons can explain these differences:
first, \citet{Belloche2019} performed several iterations of self-calibration
when reducing the interferometric data that were used here, while \citet{Li2021}
probably did not. Second, and more importantly, the determination of the
continuum level is notoriously difficult in spectra close to the confusion
limit. The automatic algorithm that we used to fit the continuum level for
each pixel of the data cube relies on the distribution of intensities of each
pixel and removes a zeroth-order baseline
\citep[see Sect. 2.2 of][]{Belloche2019}. In contrast, \citet{Li2021} fitted
first-order baselines after a manual selection of 3--5 groups of channels that
seemed to be free of line emission. These different approaches can lead in some
cases to different fitted continuum levels. This illustrates the additional
uncertainty that arises from the continuum fitting and affects the position of
the level used to assess the significance of a line detection (e.g., the
$3\sigma$ level in the figures shown in Appendix~\ref{a:spectra}).

Several portions of the ReMoCA spectrum where we find discrepancies between
the synthetic and observed spectra were not shown in Figs.~5--7 of
\citet{Li2021}, for instance the transitions at 91376~MHz, 93432~MHz, 97320~MHz,
99288~MHz, or 102568~MHz in Fig.~\ref{f:spec_c2h5conh2_ve0_p158m272},
94030~MHz or 99739~MHz in Fig.~\ref{f:spec_c2h5conh2_v30e1_p158m272}. In
addition, some of the lines of propionamide claimed by \citet{Li2021} as clearly
detected in their Fig.~5 cannot be considered as detected in our spectra, for
instance those at 92403~MHz, 92408~MHz, or 103342~MHz. Finally,
Table~\ref{t:coldens_n1s} indicates that propionamide is at least a factor 14
times less abundant than acetamide toward Sgr~B2(N1S). It would be surprising
if propionamide would be nearly as abundant as acetamide at the position
offset to the East as claimed by \citet{Li2021}. We conclude from this that
there is so far no convincing evidence for the presence of propionamide in the
interstellar medium.

\subsection{Formation mechanisms for acrylamide and related species}
\label{ss:discussion_chemistry}

In \citet{Alonso2021} we discussed the possible formation mechanisms for
propiolamide in light of its nondetection, guided by findings
from recent astrochemical kinetics models. Although presently no chemical networks
appear to include either propiolamide, acrylamide, or propionamide,
the behavior of related molecules can provide some insight into their possible
behavior. \cite{Alonso2021} proposed that propiolamide could be formed on
grain surfaces at early times in the evolution of a hot core, when the gas and
dust are cold and the dust-grain ice mantles are just beginning to form. At
that point in time, small unsaturated hydrocarbons should be relatively abundant in
the gas phase; from there, some fraction would be deposited onto the grain
surfaces, where it could react further with other species. The adsorption onto
the grains of C$_2$H in particular, or its production from other species
adsorbed from the gas phase, could lead to reaction with the radical NH$_2$CO
to form propiolamide. Alternatively, C$_2$H could react with mobile
atomic H, to produce the more stable acetylene, C$_2$H$_2$. The production on
the grains of CH$_3$ could likewise lead to a reaction with NH$_2$CO to
produce acetamide, with the main alternative being hydrogenation to CH$_4$.
Molecules formed on the grains in this way at low temperatures would
ultimately be released into the gas phase during the later hot stage of
evolution.

With this in mind, \cite{Alonso2021} used the relative solid-phase
abundances of C$_2$H$_2$ and CH$_4$ obtained in the models of \citet{Garrod21}
to infer the possible ratio of propiolamide to acetamide on the grains. Here
we extend this analysis to consider also acrylamide and propionamide, using
C$_2$H$_4$ and C$_2$H$_6$ as proxies, respectively. As before, values are taken
from the end-time of the cold collapse stage of evolution used in the models
\citep[technically, the \textit{final} model setup of][]{Garrod21}. We find
solid-phase ratios of CH$_4$ with respect to C$_2$H$_2$, C$_2$H$_4$, and
C$_2$H$_6$ of 2800, 440, and 27, respectively. This suggests that, if formed
through cold addition of C$_2$H$_3$ to NH$_2$CO on grain surfaces, acrylamide
should be around 440 times less abundant than acetamide. This value is in
nominal agreement with the upper limit values obtained for Sgr~B2(N1S) and
Sgr~B2(N2), and suggests that this molecule could have an abundance at least
an order of magnitude below those limits. We note, however, that the
ratios from the models do not consider any formation or destruction that could
occur during the hot stage, nor do they take account of differences in binding
energies or any associated variations in spatial arrangement in the gas phase,
following thermal release from the grains.

Interestingly, the inferred ratio of acetamide to propionamide is only 27,
which is indeed only a factor of two greater than the observational limit
toward Sgr~B2(N1S), while it is consistent with the nondetection in each
source. We note also that the claim of \cite{Li2021} that
propionamide is detected in Sgr~B2(N) with an abundance nearly as high as
acetamide is inconsistent with the abundance ratio that we infer for these
species from the models.

As noted by \citet{Alonso2021}, propiolamide (formed in whatever fashion) could
be further hydrogenated on the dust grains all the way to propionamide. Two
necessary steps in this process would be the production, and then destruction,
of acrylamide, through the addition of atomic H. However, abstraction of H
from any of these species or their intermediates through alternative reaction
pathways with mobile atomic H could result in conversion in the other
direction. Depending on the balance of activation energy barriers and
branching ratios, this could therefore lead to some stable ratio between
acrylamide and the other two. In fact, something like this situation occurs in
the astrochemical models for the species C$_2$H$_2$, C$_2$H$_4$, and C$_2$H$_6$
themselves. One should therefore expect that the ratios derived above would
already reflect this sort of relationship (any dissimilarities in activation
energy barriers notwithstanding).

Alternative mechanisms for producing propiolamide, acrylamide, and
propionamide, such as NH$_2$ addition to the radical C$_2$H$_3$CO, are more
difficult to assess based on inference from the models, but they also appear
plausible. Another alternative mechanism for propionamide formation would be a
more direct conversion of acetamide, via the abstraction of an H atom from the
methyl group (by H or some reactive radical such as OH), followed by the
addition of another methyl group. \citet{Garrod21} also invoked the reaction
of methylene (CH$_2$) with methanol (CH$_3$OH) on the grains, to produce
ethanol (C$_2$H$_5$OH) as well as other products. Methylene reactions might
provide a route to lengthen alkyl groups in other molecules, including
acetamide. This could suggest perhaps an even greater abundance for
propionamide, by some unknown degree.

A more rugged test of these ideas would involve the development of a
self-consistent network for all of the above species, as well as for related
alkyl-group bearing molecules whose production may be competitive with the
others. Application of the network to the full hot-core physical treatment
would then be possible.

\section{Conclusions}
\textbf{\label{s:conclusions}}

In the present work we conducted a detailed rotational study of acrylamide, a peptide bond bearing species, from 75 to 480~GHz. More than 6000 new rotational lines for the ground state and two excited vibrational states were measured and assigned for the most stable \textit{syn} conformer. In addition to this, over 2500 lines were analyzed for the less stable \textit{skew} conformer. The ground vibrational state of this conformer was observed in the form of a doublet. We interpreted this observation as a consequence of a double-minimum potential function with a small barrier between two equivalent \textit{skew} structures.
A comprehensive spectral analysis provided accurate frequency predictions to search for acrylamide in Sgr B2(N). Additionally, emission lines from the related molecule propionamide were searched for in the same source. The main conclusions of these searches are following:
\begin{enumerate}
\item Acrylamide was not detected toward the hot molecule cores Sgr~B2(N1S)
and Sgr~B2(N2) with ALMA. The upper limits derived for its column density
imply that it is at least 26 and 6 times less abundant than acetamide toward
these sources, respectively.
\item Propionamide was not detected toward Sgr~B2(N1S) and Sgr~B2(N2) either.
It is at least 14 and 3 times less abundant than acetamide toward these sources,
respectively.
\item We do not confirm either the tentative detection of propionamide
recently reported in the literature toward a position offset to the East of
Sgr~B2(N1). We conclude from this that
there is so far no convincing evidence for the presence of propionamide in the
interstellar medium.
\item While acrylamide is not yet present in any astrochemical kinetics models,
comparison with model results for related species suggests that acrylamide
may be a few hundred times less abundant than acetamide, corresponding to a
value at least an order of magnitude lower than the observational upper
limits. Propionamide may be as little as a factor of two less abundant than
the upper limit toward Sgr B2(N1S).
\end{enumerate}

\begin{acknowledgements}

This work has been supported by the Czech Science Foundation (GACR, grant 19-25116Y). The funding is gratefully acknowledged.
We thank Roman Motiyenko and Laurent Margul\`es for sending us spectroscopic
predictions for propionamide in electronic format. This paper makes use of the
following ALMA data: ADS/JAO.ALMA\#2016.1.00074.S.
ALMA is a partnership of ESO (representing its member states), NSF (USA), and
NINS (Japan), together with NRC (Canada), NSC and ASIAA (Taiwan), and KASI
(Republic of Korea), in cooperation with the Republic of Chile. The Joint ALMA
Observatory is operated by ESO, AUI/NRAO, and NAOJ. The interferometric data
are available in the ALMA archive at https://almascience.eso.org/aq/.
Part of this work has been carried out within the Collaborative
Research Centre 956, sub-project B3, funded by the Deutsche
Forschungsgemeinschaft (DFG) -- project ID 184018867.
RTG acknowledges support from the National Science Foundation (grant No.
AST 19-06489).

\end{acknowledgements}



\bibliography{library}



\begin{appendix}

\section{Complementary Tables}

Table \ref{transitions-syn} lists the measured transitions of \textit{syn} acrylamide.
Table \ref{vib-modes} lists the frequencies of normal vibrational modes of \textit{syn} and \textit{skew} acrylamide used to calculate the vibrational partition function.
Table \ref{Tcomparison} lists a comparison of the ground state spectroscopic constants of \textit{syn} acrylamide obtained from different data sets.
Table \ref{transitions-skew} lists the measured transitions of \textit{skew} acrylamide.

\begin{table*}[!h]
\caption{List of the measured transitions of \textit{syn} acrylamide.}
\label{transitions-syn}
\begin{center}
\begin{footnotesize}
\setlength{\tabcolsep}{4pt}
\begin{tabular}{l r r r r r r r r c c r r r}
\hline\hline
Vib. state & $J'$ & $K_{a}'$  & $K_{c}'$ & $J''$ & $K_{a}''$ & $K_{c}''$ & $\nu_{\text{obs}}$ (MHz) \tablefootmark{a} & $\nu_{\text{obs}}-\nu_{\text{calc}}$ (MHz) \tablefootmark{b} & $u_{\text{obs}}$ (MHz) \tablefootmark{c}  & $(\nu_{\text{obs}}-\nu_{\text{calc}})_{\text{blends}}$ (MHz) \tablefootmark{d} & Weight \tablefootmark{e} & Notes\tablefootmark{f}\\
\hline
G.S.         & 15 & 4 & 11 &  15 & 3 & 12 &   38737.1700 &   0.0844  &  0.100 &           &       &(1) \\
G.S.         & 13 & 2 & 12 &  12 & 1 & 11 &   88541.9842 &   0.0120  &  0.020 &           &       &(2) \\
G.S.         & 16 & 1 & 16 &  15 & 0 & 15 &  100009.6024 &   0.0088  &  0.020 &           &       &(2) \\
G.S.         & 32 & 3 & 30 &  31 & 2 & 29 &  209078.0213 & --0.0381  &  0.020 &  0.0126   & 0.50  &(2) \\
G.S.         & 32 & 2 & 30 &  31 & 3 & 29 &  209078.0213 &   0.0632  &  0.020 &  0.0126   & 0.50  &(2) \\
$v_{24}=1$   & 14 & 2 & 12 &  13 & 3 & 11 &   96383.9981 & --0.0071  &  0.020 &           &       &(2) \\
$v_{24}=1$   & 41 & 3 & 38 &  40 & 4 & 37 &  270039.3106 &   0.0019  &  0.020 & --0.0009  & 0.50  &(2) \\
$v_{24}=1$   & 41 & 4 & 38 &  40 & 3 & 37 &  270039.3106 & --0.0038  &  0.020 & --0.0009  & 0.50  &(2) \\
$v_{24}=2$   & 28 & 5 & 24 &  27 & 4 & 23 &  197984.2945 &   0.0226  &  0.020 &           &       &(2) \\
$v_{24}=2$   & 65 & 9 & 56 &  64 &10 & 55 &  452320.1788 &   0.0130  &  0.020 & --0.0192  & 0.50  &(2) \\
$v_{24}=2$   & 65 &10 & 56 &  64 & 9 & 55 &  452320.1788 & --0.0513  &  0.020 & --0.0192  & 0.50  &(2)\\
\hline
\end{tabular}
\end{footnotesize}
\end{center}
\tablefoot{
\tablefoottext{a}{Observed frequency.}
\tablefoottext{b}{Observed minus calculated frequency.}
\tablefoottext{c}{Uncertainty of the observed frequency.}
\tablefoottext{d}{Observed minus calculated frequency for blends.}
\tablefoottext{e}{Intensity weighting factor for blended transitions.}
\tablefoottext{f}{Source of the data: (1) \cite{Marstokk2000}, (2) This work.}
This table is available in its entirety in electronic form at the CDS via anonymous ftp to cdsarc.u-strasbg.fr (130.79.128.5) or via
http://cdsweb.u-strasbg.fr/cgi-bin/qcat?J/A+A/. A portion is shown here for guidance regarding its form and content.}
\end{table*}

\begin{table*}
\caption{Frequencies of normal vibrational modes of the two conformers of acrylamide used to calculate their vibrational partition function.}
\label{vib-modes}
\begin{center}
\begin{footnotesize}
\setlength{\tabcolsep}{5pt}
\begin{tabular}{r r r r r r }
\hline\hline
     & \multicolumn{2}{c}{\textit{Syn} conformer} & &  \multicolumn{2}{c}{\textit{Skew} conformer} \\
\cline{2-3}
\cline{5-6}
Mode & Frequency (cm$^{-1}$) & Symmetry & & Frequency (cm$^{-1}$) & Symmetry \\
\hline
 1    & 3727.6   &  $A^\prime$                          &   & 3749.7    & $A$    \\
 2    & 3593.3   &  $A^\prime$                          &   & 3584.2    & $A$    \\
 3    & 3236.1   &  $A^\prime$                          &   & 3219.3    & $A$    \\
 4    & 3151.1   &  $A^\prime$                          &   & 3178.8    & $A$    \\
 5    & 3140.0   &  $A^\prime$                          &   & 3137.9    & $A$    \\
 6    & 1764.3   &  $A^\prime$                          &   & 1758.6    & $A$    \\
 7    & 1689.6   &  $A^\prime$                          &   & 1684.8    & $A$    \\
 8    & 1617.3   &  $A^\prime$                          &   & 1617.6    & $A$    \\
 9    & 1443.8   &  $A^\prime$                          &   & 1455.7    & $A$    \\
10    & 1356.3   &  $A^\prime$                          &   & 1359.9    & $A$    \\
11    & 1290.0   &  $A^\prime$                          &   & 1311.6    & $A$    \\
12    & 1113.2   &  $A^\prime$                          &   & 1116.6    & $A$    \\
13    & 1037.9   &  $A^\prime$                          &   & 1043.9    & $A$    \\
14    &  811.5   &  $A^\prime$                          &   & 1040.3    & $A$    \\
15    &  616.8   &  $A^\prime$                          &   &  986.4    & $A$    \\
16    &  466.2   &  $A^\prime$                          &   &  824.1    & $A$    \\
17    &  276.9   &  $A^\prime$                          &   &  819.0    & $A$    \\
18    & 1018.8   &  $A^{\prime\prime}$                  &   &  592.9    & $A$    \\
19    & 1008.6   &  $A^{\prime\prime}$                  &   &  561.3    & $A$    \\
20    &  821.2   &  $A^{\prime\prime}$                  &   &  525.1    & $A$    \\
21    &  616.9   &  $A^{\prime\prime}$                  &   &  432.0    & $A$    \\
22    &  469.2   &  $A^{\prime\prime}$                  &   &  353.8    & $A$    \\
23    &  262.8 \tablefootmark{a}   &  $A^{\prime\prime}$&   &  277.6    & $A$    \\
24    &   90 \tablefootmark{b}     &  $A^{\prime\prime}$&   &  99.9     & $A$    \\
\hline
\end{tabular}
\end{footnotesize}
\end{center}
\tablefoot{
\tablefoottext{a}{Taken from the IR measurements from \cite{Kydd1980}}.
\tablefoottext{b}{From relative intensity measurements from \cite{Marstokk2000}. The stated uncertainty is 10 cm$^{-1}$}.
The rest of the vibrational modes is taken from the theoretical calculations at the B3LYP/cc-pVTZ level of theory performed by \cite{Marstokk2000}.}
\end{table*}

\begin{table*}[ht]
\caption{Comparison of the ground state spectroscopic constants of \textit{syn} acrylamide obtained from different data sets ($A$-reduction, I$^{\text{r}}$-representation).}
\label{Tcomparison}
\begin{center}
\begin{footnotesize}
\setlength{\tabcolsep}{5.0pt}
\begin{tabular}{ l r r  r r }
\hline\hline
 & Data set I\tablefootmark{a} & Data set II\tablefootmark{b}  & Data set III\tablefootmark{c}  & Data set IV\tablefootmark{d} \\
\hline
$A                   $  /               MHz    &  10732.8296 (34)\tablefootmark{e}    &   10732.8220 (33)        &  10732.819329 (66)    &   10732.819331 (65)  \\
$B                   $  /               MHz    &   4218.7012 (13)    &    4218.6933 (11)        &   4218.690258 (41)    &    4218.690256 (41)  \\
$C                   $  /               MHz    &   3030.7434 (13)    &    3030.7494 (23)        &   3030.752982 (33)    &    3030.752979 (33)  \\
$\Delta_{J}          $  /               kHz    &      0.7043 (33)    &       0.7396 (61)        &      0.786817 (13)    &       0.786816 (13)  \\
$\Delta_{JK}         $  /               kHz    &      3.370 (22)     &       3.467 (44)         &      3.755001 (41)    &       3.755014 (41)  \\
$\Delta_{K}          $  /               kHz    &      5.403 (8)      &       5.458 (17)         &      5.49543 (11)     &       5.49540 (11)   \\
$\delta_{J}          $  /               kHz    &      0.2417 (7)     &       0.2249 (11)        &      0.2227674 (57)   &       0.2227676 (56) \\
$\delta_{K}          $  /               kHz    &      3.20 (4)       &       3.315 (81)         &      3.361824 (72)    &       3.361833 (72)  \\
$\Phi_{J}            $  /               mHz    &     --66 (6)        &     ...\tablefootmark{f}                &       0.1763 (16)     &       0.1763 (16)    \\
$\Phi_{JK}           $  /               mHz    &    --222 (34)       &     ...\tablefootmark{f}                &       1.357 (16)      &       1.361 (16)     \\
$\Phi_{KJ}           $  /               mHz    &    --1540 (130)     &    --23.17(58)           &    --30.829 (56)      &    --30.836 (56)     \\
$\Phi_{K}            $  /               mHz    &     570 (40)        &      54.8(10)            &      55.102 (77)      &      55.087 (76)     \\
$\phi_{J}            $  /               mHz    &    1.70 (34)        &     ...\tablefootmark{f}                &       0.09113 (78)    &       0.09118 (77)   \\
$\phi_{JK}           $  /               mHz    &    --600 (50)       &     ...\tablefootmark{f}                &       1.102 (16)      &       1.104 (16)     \\
$\phi_{K}            $  /               mHz    &     1980 (160)      &     ...\tablefootmark{f}                &      40.42 (11)       &      40.43 (11)      \\
$J_{\text{min}}/J_{\text{max}}$                &    1 / 75           &     4 / 75               &       3 / 78              &    3 / 78            \\
$K_{a}^{\text{min}}/K_{a}^{\text{max}}$        &    0 / 37           &     0 / 37               &       0 / 33              &    0 / 37            \\
$N$\tablefootmark{g}                           &   354               &    172                   &       2538              &    2710              \\
$\sigma_{\text{fit}}$\tablefootmark{h}/ MHz    &   0.081             &    0.065                 &       0.019                  &    0.028             \\
\hline
\end{tabular}
\end{footnotesize}
\end{center}
\tablefoot{
\tablefoottext{a}{Whole data set from \cite{Marstokk2000}.}
\tablefoottext{b}{Data from \cite{Marstokk2000} when the misassigned lines and hyperfine-split lines were removed.}
\tablefoottext{c}{Data solely from the present work.}
\tablefoottext{d}{A global fit including the data from the present work and those from the second column of this table.}
\tablefoottext{e}{The numbers in parentheses are the parameter uncertainties in units of the last decimal digit. SPFIT/SPCAT program package \citep{Pickett1991} was used for the analysis.}
\tablefoottext{f}{Fixed to zero.}
\tablefoottext{g}{Number of distinct frequency lines in the fit.}
\tablefoottext{h}{Root mean square deviation of the fit.}}
\end{table*}

\begin{table*}[!h]
\caption{List of the measured transitions of \textit{skew} acrylamide.}
\label{transitions-skew}
\begin{center}
\begin{footnotesize}
\setlength{\tabcolsep}{4pt}
\begin{tabular}{l r r r r r r r r r c c r r r}
\hline\hline
 $J'$ & $K_{a}'$  & $K_{c}'$ & $v'$ \tablefootmark{a} & $J''$ & $K_{a}''$ & $K_{c}''$ & $v''$ \tablefootmark{a} & $\nu_{\text{obs}}$ (MHz) \tablefootmark{b} & $\nu_{\text{obs}}-\nu_{\text{calc}}$ (MHz) \tablefootmark{c} & $u_{\text{obs}}$ (MHz) \tablefootmark{d}  & $(\nu_{\text{obs}}-\nu_{\text{calc}})_{\text{blends}}$ (MHz) \tablefootmark{e} & Weight \tablefootmark{f} & Notes\tablefootmark{g}\\
\hline
  8 &  5 &  3 &  0 &    7 &  5 &  2 &  0   &    59501.6900 & --0.0654 &  0.100 &           &        &(1)\\
 15 &  1 & 15 &  0 &   14 &  1 & 14 &  0   &    94032.4063 & --0.0025 &  0.050 &           &        &(2)\\
 26 &  1 & 25 &  0 &   25 &  2 & 24 &  0   &   166718.5703 &   0.0045 &  0.030 & --0.0114  &  0.29  &(2)\\
 26 &  2 & 25 &  0 &   25 &  2 & 24 &  0   &   166718.5703 & --0.0053 &  0.030 & --0.0114  &  0.21  &(2)\\
 26 &  2 & 25 &  0 &   25 &  1 & 24 &  0   &   166718.5703 & --0.0273 &  0.030 & --0.0114  &  0.29  &(2)\\
 26 &  1 & 25 &  0 &   25 &  1 & 24 &  0   &   166718.5703 & --0.0174 &  0.030 & --0.0114  &  0.21  &(2)\\
 16 &  0 & 16 &  1 &   15 &  1 & 15 &  1   &   100553.9229 &   0.0256 &  0.050 &           &        &(2)\\
 26 &  7 & 20 &  1 &   25 &  7 & 19 &  1   &   196306.1152 &   0.0069 &  0.030 &           &        &(2)\\
 22 & 17 &  5 &  1 &   21 & 16 &  6 &  1   &   370327.0981 & --0.0034 &  0.030 & --0.0034  &  0.50  &(2)\\
 22 & 17 &  6 &  1 &   21 & 16 &  5 &  1   &   370327.0981 & --0.0034 &  0.030 & --0.0034  &  0.50  &(2)\\
\hline
\end{tabular}
\end{footnotesize}
\end{center}
\tablefoot{
\tablefoottext{a}{$v=0$ corresponds to $0^+$ state and $v=1$ to $0^-$ state.}
\tablefoottext{b}{Observed frequency.}
\tablefoottext{c}{Observed minus calculated frequency.}
\tablefoottext{d}{Uncertainty of the observed frequency.}
\tablefoottext{e}{Observed minus calculated frequency for blends.}
\tablefoottext{f}{Intensity weighting factor for blended transitions.}
\tablefoottext{g}{Source of the data: (1) \cite{Marstokk2000}, (2) This work.}
This table is available in its entirety in electronic form at the CDS via anonymous ftp to cdsarc.u-strasbg.fr (130.79.128.5) or via
http://cdsweb.u-strasbg.fr/cgi-bin/qcat?J/A+A/. A portion is shown here for guidance regarding its form and content.}
\end{table*}

\section{Complementary figures: Astronomical spectra}
\label{a:spectra}

\begin{figure*}[!t]
\centerline{\resizebox{0.88\hsize}{!}{\includegraphics[angle=0]{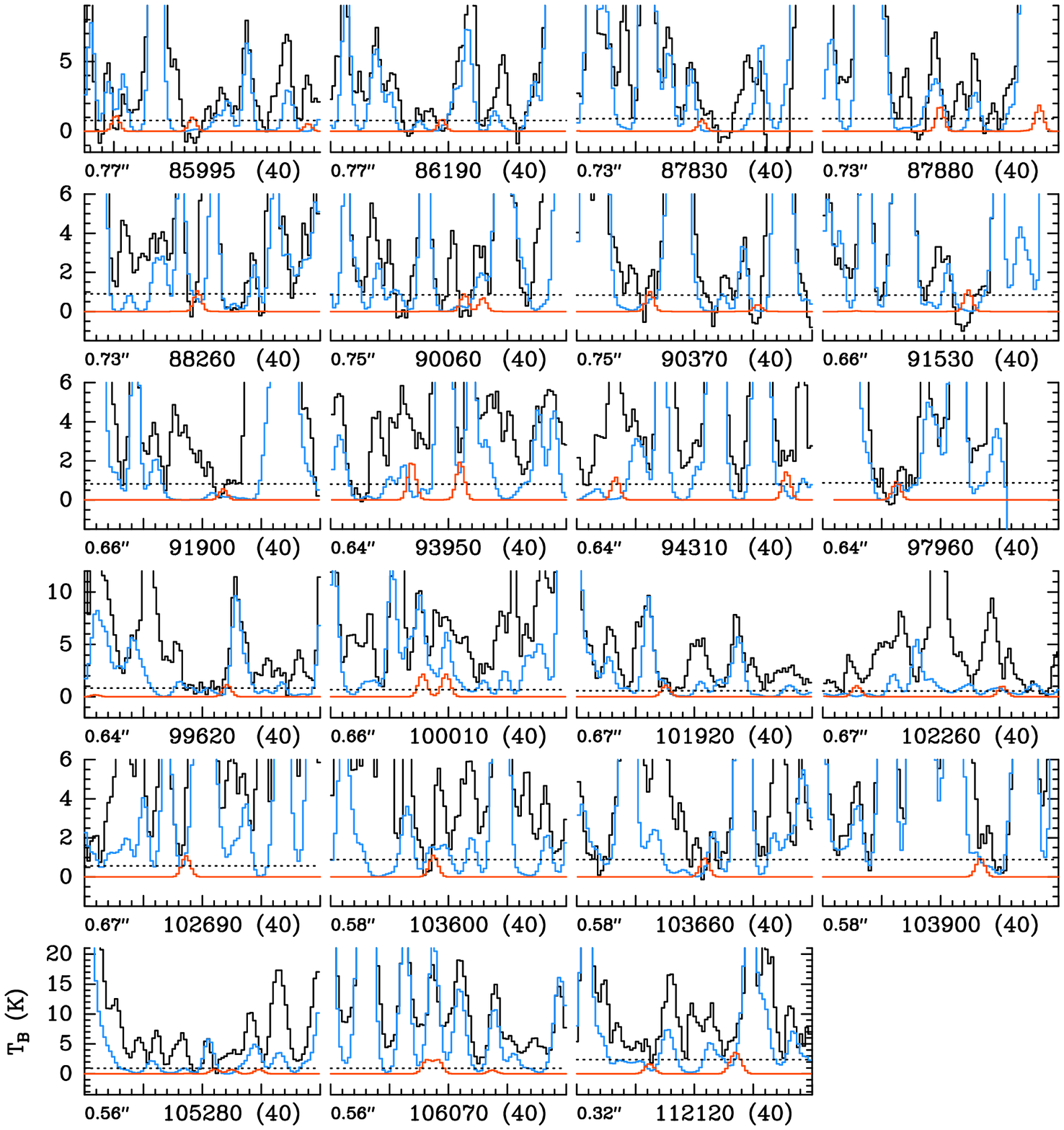}}}
\caption{Selection of transitions of acrylamide,
\textit{syn}-C$_2$H$_3$C(O)NH$_2$, $\varv = 0$ covered by the ReMoCA survey. The
synthetic spectrum of
\textit{syn}-C$_2$H$_3$C(O)NH$_2$, $\varv = 0$ used to derive the upper limit to
its column density is displayed in red and overlaid on the observed spectrum
of Sgr~B2(N1S) shown in black. The blue synthetic spectrum contains the
contributions from all molecules identified in our survey so far, but not from
the species shown in red. The central frequency and width are indicated in MHz
below each panel. The angular resolution (HPBW) is also indicated. The
$y$-axis is labeled in brightness temperature units (K). The dotted line
indicates the $3\sigma$ noise level.}
\label{f:spec_c2h3conh2-s_ve0_n1s}
\end{figure*}

\begin{figure*}[!t]
\centerline{\resizebox{0.695\hsize}{!}{\includegraphics[angle=0]{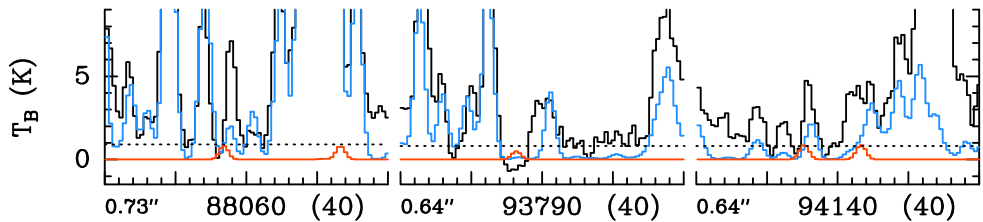}}}
\caption{Same as Fig.~\ref{f:spec_c2h3conh2-s_ve0_n1s} but for acrylamide,
\textit{syn}-C$_2$H$_3$C(O)NH$_2$, $\varv_{24} = 1$ toward Sgr~B2(N1S).}
\label{f:spec_c2h3conh2-s_v24e1_n1s}
\end{figure*}

\begin{figure*}[!t]
\centerline{\resizebox{0.88\hsize}{!}{\includegraphics[angle=0]{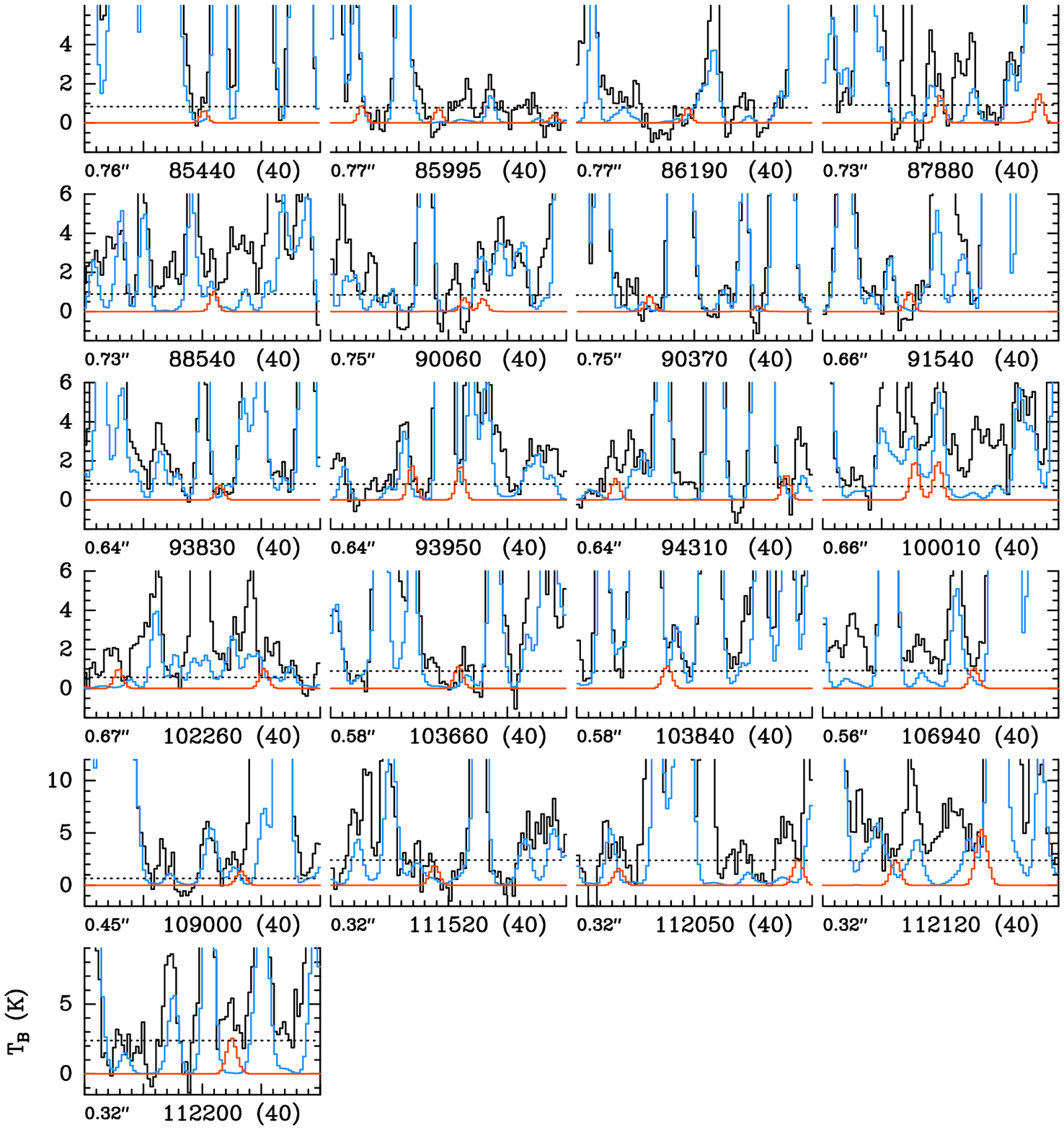}}}
\caption{Same as Fig.~\ref{f:spec_c2h3conh2-s_ve0_n1s} but for acrylamide,
\textit{syn}-C$_2$H$_3$C(O)NH$_2$, $\varv = 0$ toward Sgr~B2(N2).}
\label{f:spec_c2h3conh2-s_ve0_n2}
\end{figure*}

\begin{figure*}[!t]
\centerline{\resizebox{0.47\hsize}{!}{\includegraphics[angle=0]{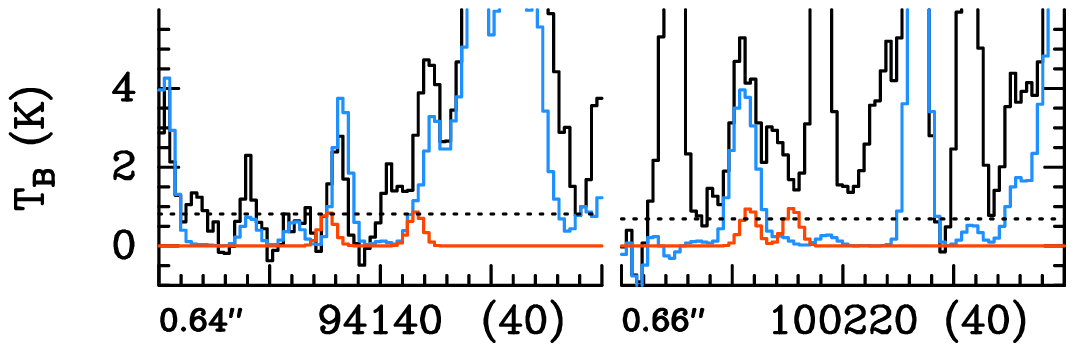}}}
\caption{Same as Fig.~\ref{f:spec_c2h3conh2-s_ve0_n1s} but for acrylamide,
\textit{syn}-C$_2$H$_3$C(O)NH$_2$, $\varv_{24} = 1$ toward Sgr~B2(N2).}
\label{f:spec_c2h3conh2-s_v24e1_n2}
\end{figure*}

\begin{figure*}[!t]
\centerline{\resizebox{0.88\hsize}{!}{\includegraphics[angle=0]{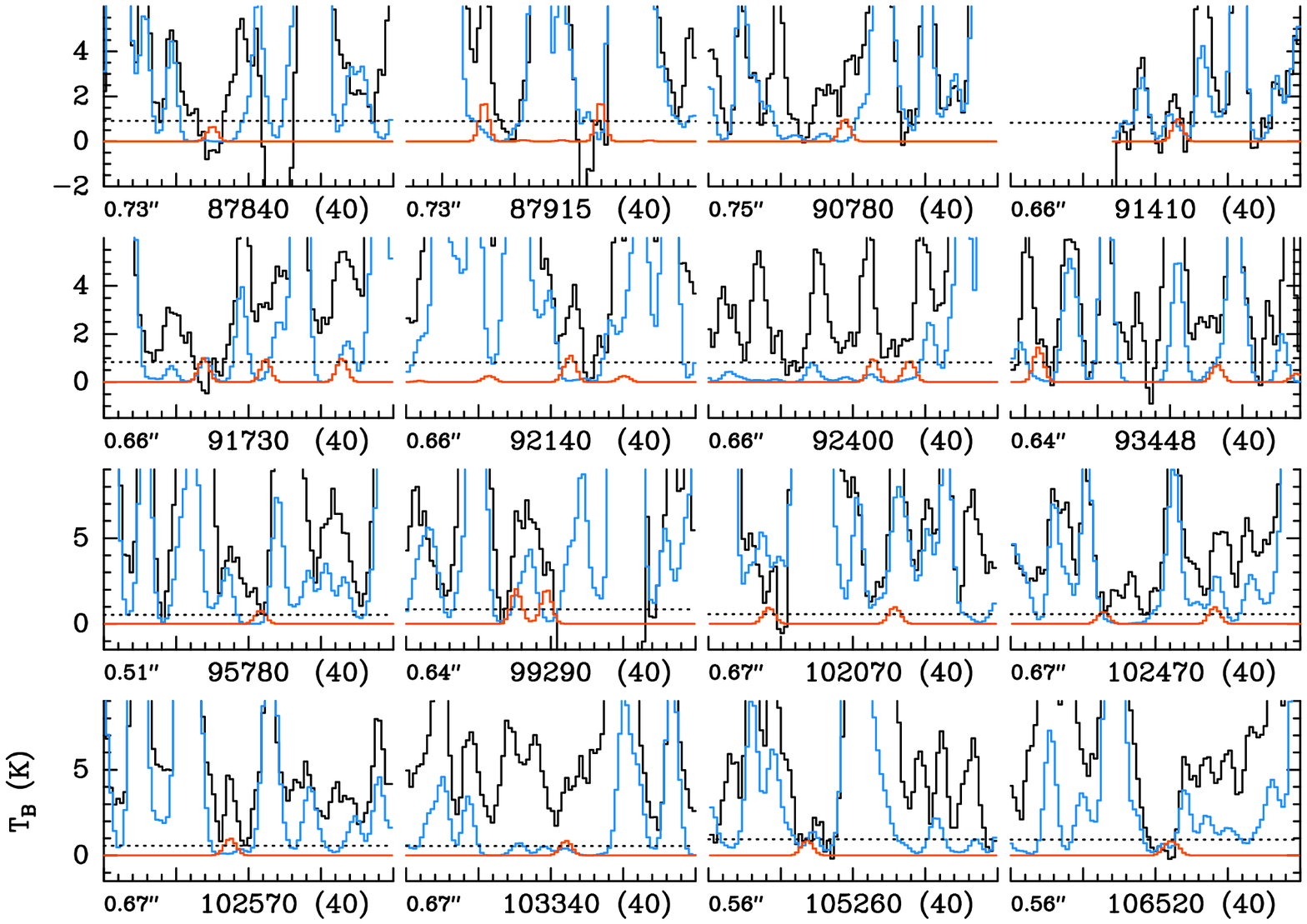}}}
\caption{Same as Fig.~\ref{f:spec_c2h3conh2-s_ve0_n1s} but for propionamide,
C$_2$H$_5$C(O)NH$_2$, $\varv = 0$ toward Sgr~B2(N1S).}
\label{f:spec_c2h5conh2_ve0_n1s}
\end{figure*}

\begin{figure*}[!t]
\centerline{\resizebox{0.88\hsize}{!}{\includegraphics[angle=0]{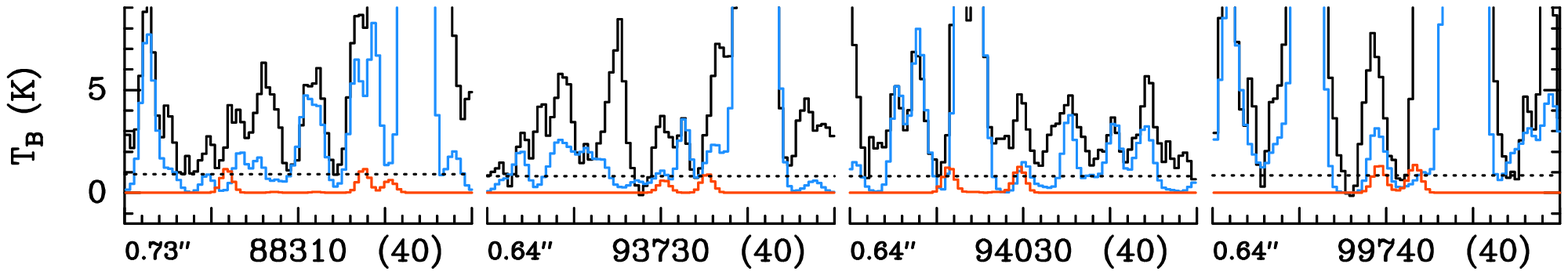}}}
\caption{Same as Fig.~\ref{f:spec_c2h3conh2-s_ve0_n1s} but for propionamide,
C$_2$H$_5$C(O)NH$_2$, $\varv_{30} = 1$ toward Sgr~B2(N1S).}
\label{f:spec_c2h5conh2_v30e1_n1s}
\end{figure*}

\begin{figure*}[!t]
\centerline{\resizebox{0.88\hsize}{!}{\includegraphics[angle=0]{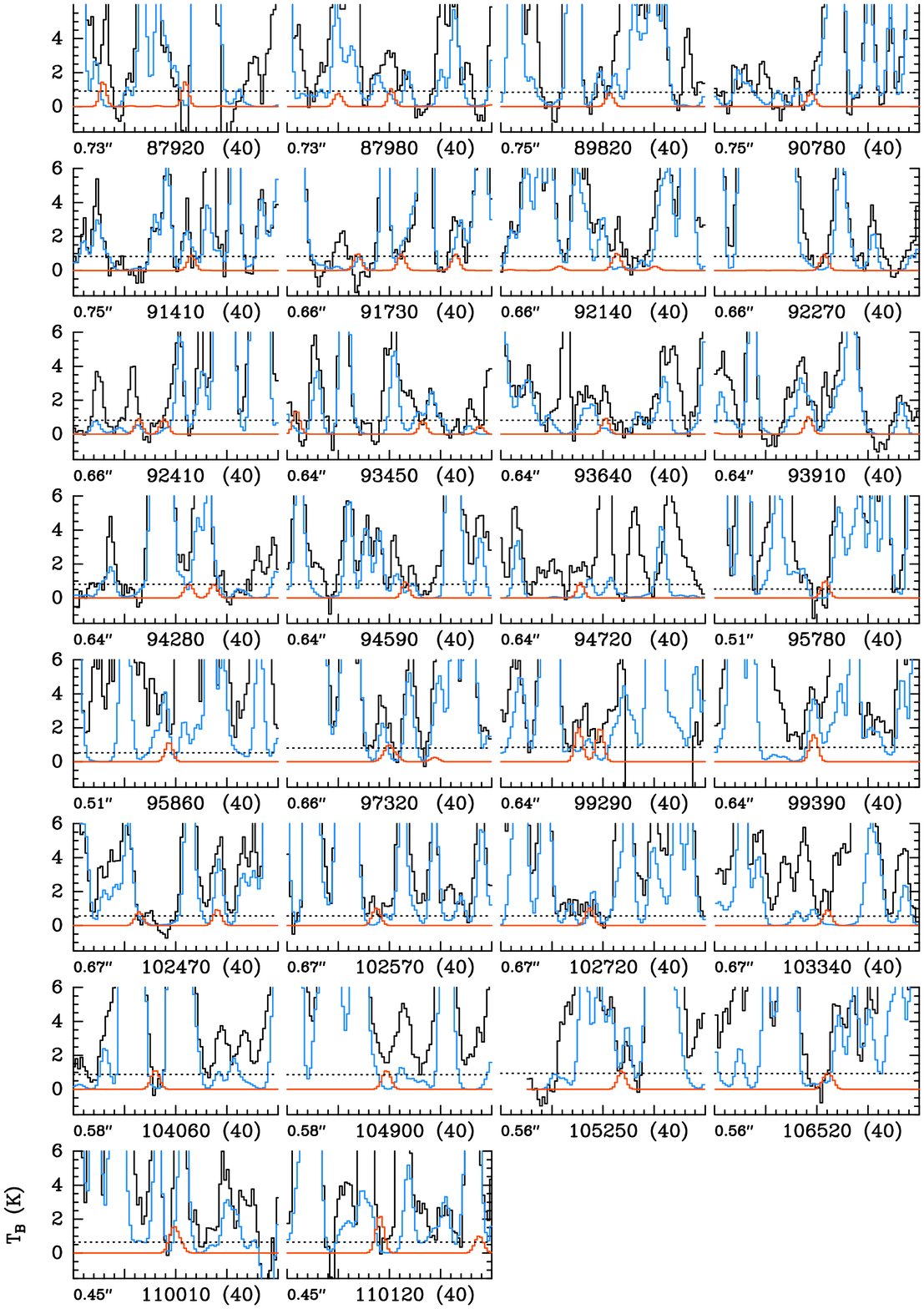}}}
\caption{Same as Fig.~\ref{f:spec_c2h3conh2-s_ve0_n1s} but for propionamide,
C$_2$H$_5$C(O)NH$_2$, $\varv=0$ toward Sgr~B2(N2).}
\label{f:spec_c2h5conh2_ve0_n2}
\end{figure*}

\begin{figure*}[!t]
\centerline{\resizebox{0.88\hsize}{!}{\includegraphics[angle=0]{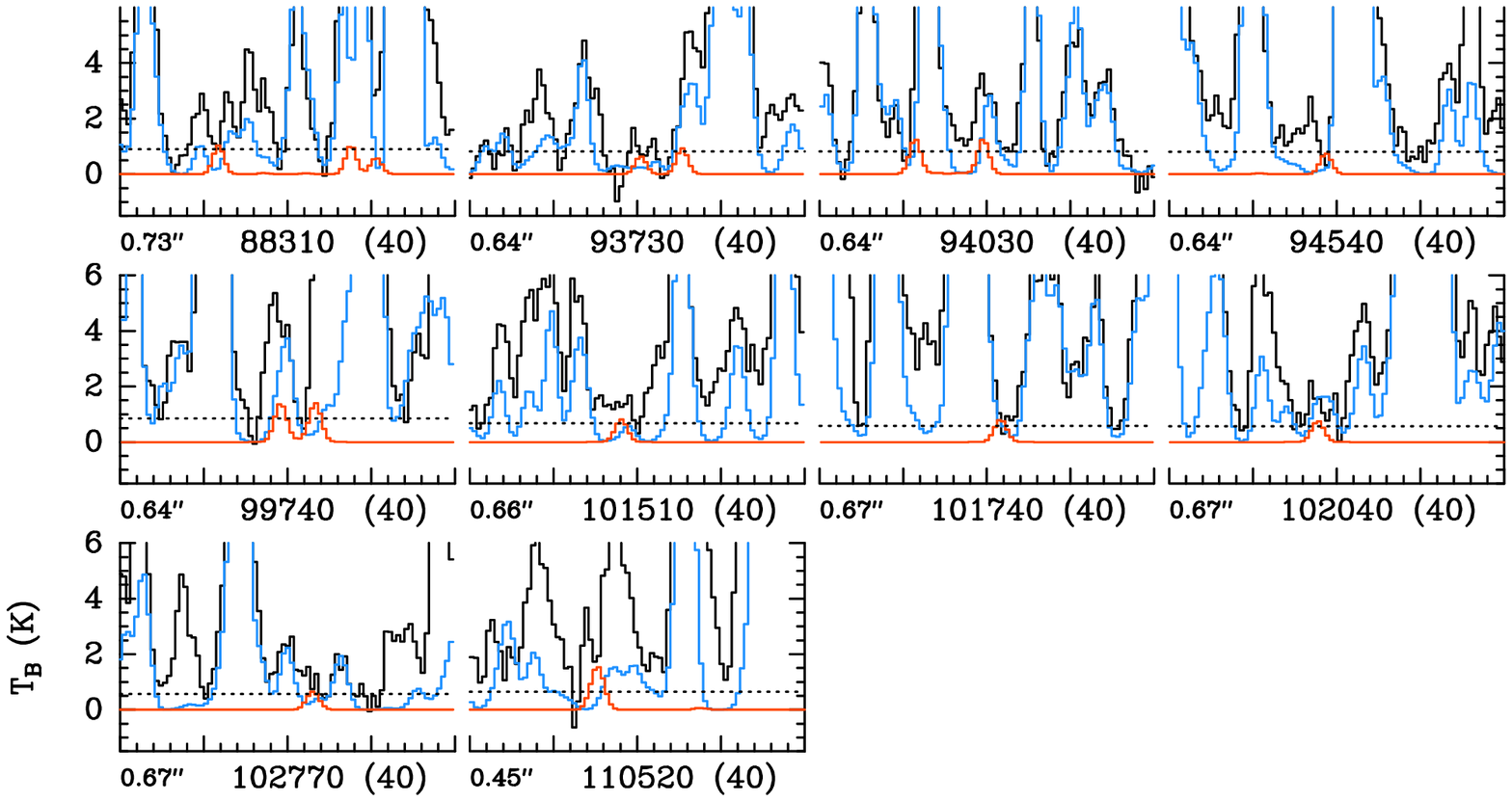}}}
\caption{Same as Fig.~\ref{f:spec_c2h3conh2-s_ve0_n1s} but for propionamide,
C$_2$H$_5$C(O)NH$_2$, $\varv_{30}=1$ toward Sgr~B2(N2).}
\label{f:spec_c2h5conh2_v30e1_n2}
\end{figure*}

Figures~\ref{f:spec_c2h3conh2-s_ve0_n1s}--\ref{f:spec_c2h3conh2-s_v24e1_n2}
illustrate the nondetection of acrylamide in its ground state, $\varv = 0$, and
its first vibrationally excited state, $\varv_{24}=1$, toward Sgr~B2(N1S) and
Sgr~B2(N2).
Figures~\ref{f:spec_c2h5conh2_ve0_n1s}--\ref{f:spec_c2h5conh2_v30e1_n2}
illustrate the nondetection of propionamide in its ground state, $\varv = 0$,
and its first vibrationally excited state, $\varv_{30}=1$, toward Sgr~B2(N1S)
and Sgr~B2(N2).
Figures~\ref{f:spec_c2h5conh2_ve0_p158m272} and
\ref{f:spec_c2h5conh2_v30e1_p158m272} illustrate the nondetection of
propionamide in its ground state, $\varv = 0$, and its first vibrationally
excited state, $\varv_{30}=1$, toward  the equatorial offset position
($1.58\arcsec$, $-2.72\arcsec$) relative to the phase center.

\begin{figure*}[!t]
\centerline{\resizebox{0.88\hsize}{!}{\includegraphics[angle=0]{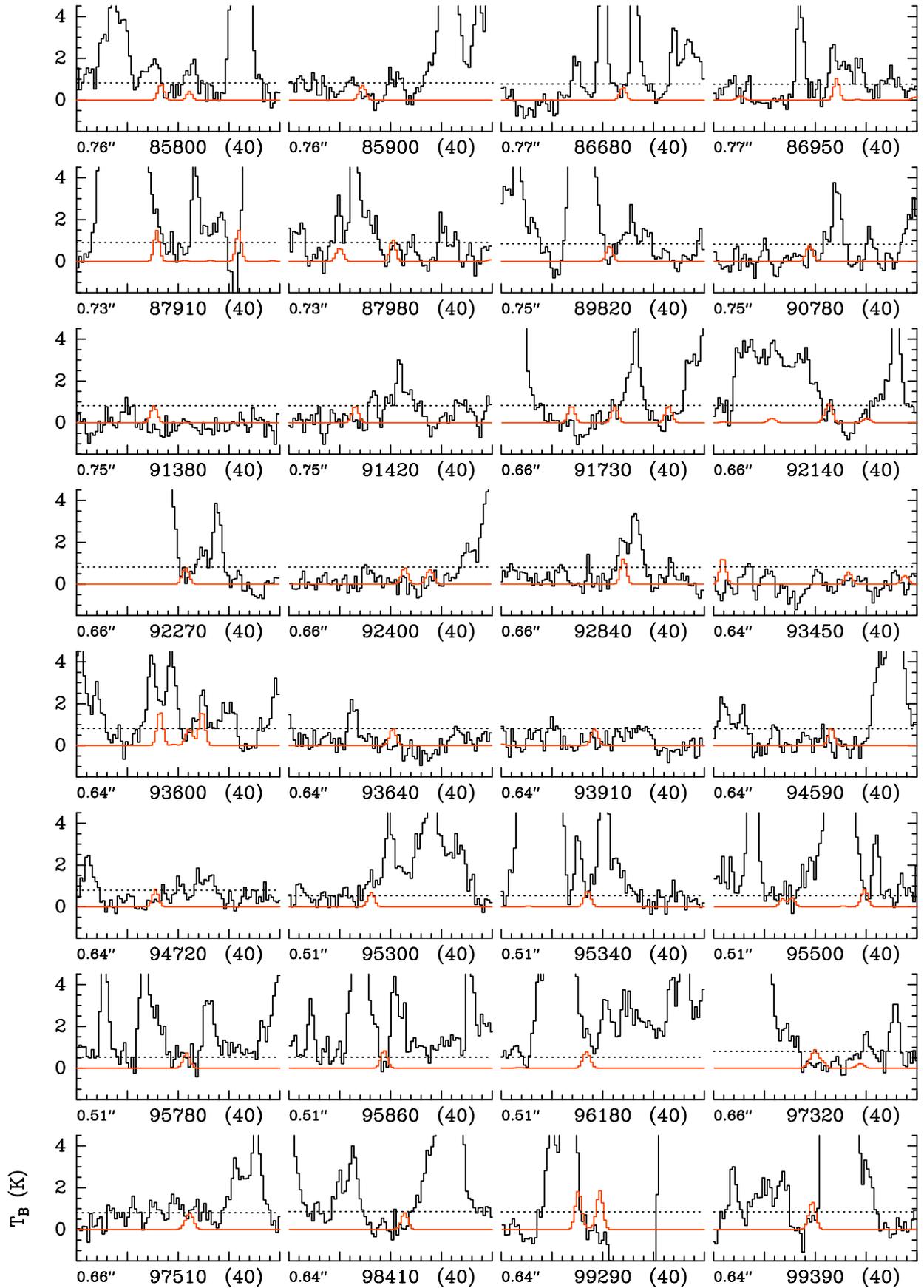}}}
\caption{Selection of transitions of propionamide, C$_2$H$_5$C(O)NH$_2$,
$\varv=0$ covered by the ReMoCA survey. The synthetic spectrum of
C$_2$H$_5$C(O)NH$_2$, $\varv=0$ used to derive the upper limit to its column
density is displayed in red and overlaid on the black spectrum observed toward
the equatorial offset position ($1.58\arcsec$, $-2.72\arcsec$) relative to the
phase center. The central frequency and width are indicated in MHz below each
panel. The angular resolution (HPBW) is also indicated. The $y$-axis is
labeled in brightness temperature units (K). The dotted line indicates the
$3\sigma$ noise level.}
\label{f:spec_c2h5conh2_ve0_p158m272}
\end{figure*}

\begin{figure*}[!t]
\addtocounter{figure}{-1}
\centerline{\resizebox{0.88\hsize}{!}{\includegraphics[angle=0]{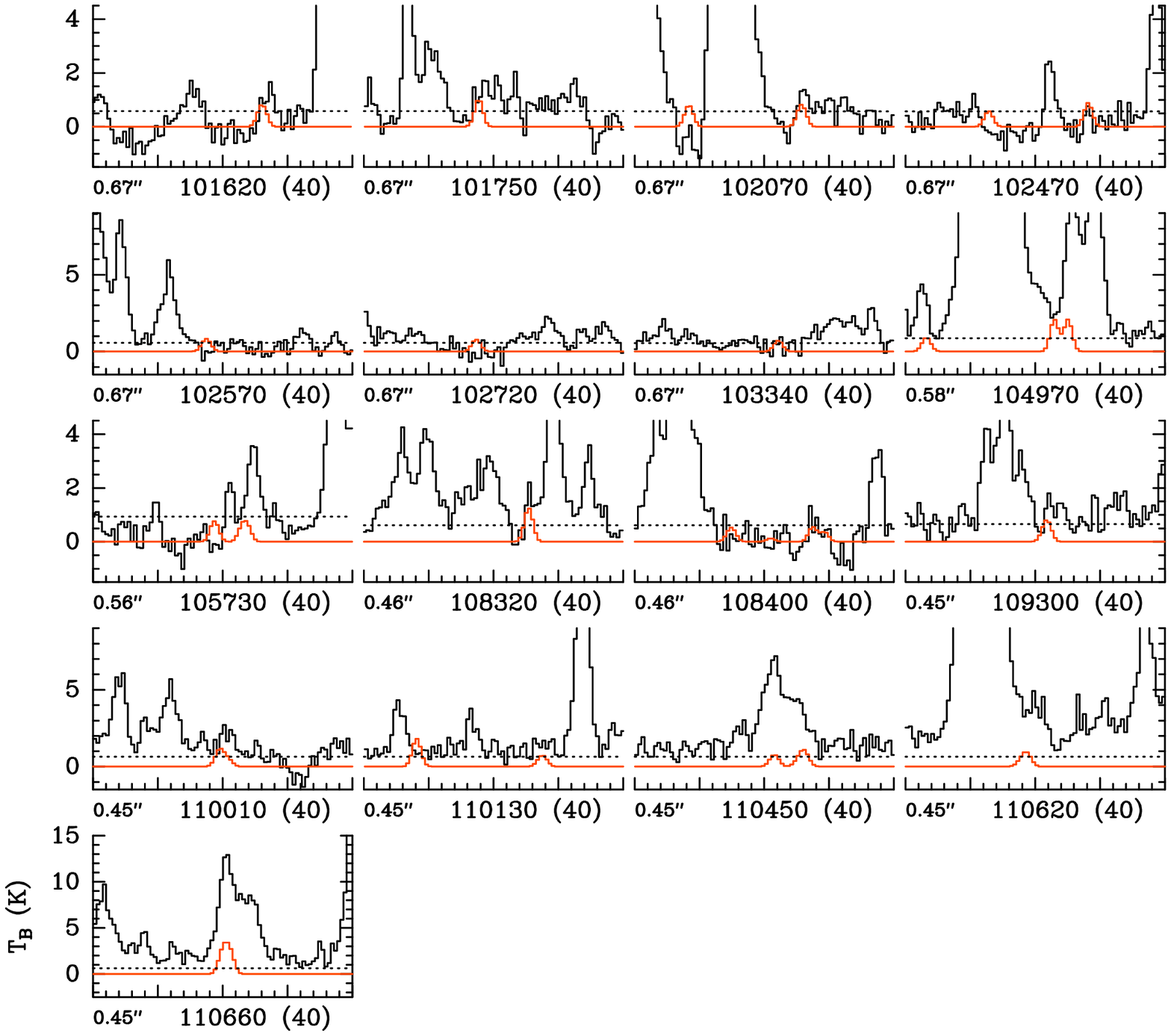}}}
\caption{continued.}
\end{figure*}

\begin{figure*}[!t]
\centerline{\resizebox{0.88\hsize}{!}{\includegraphics[angle=0]{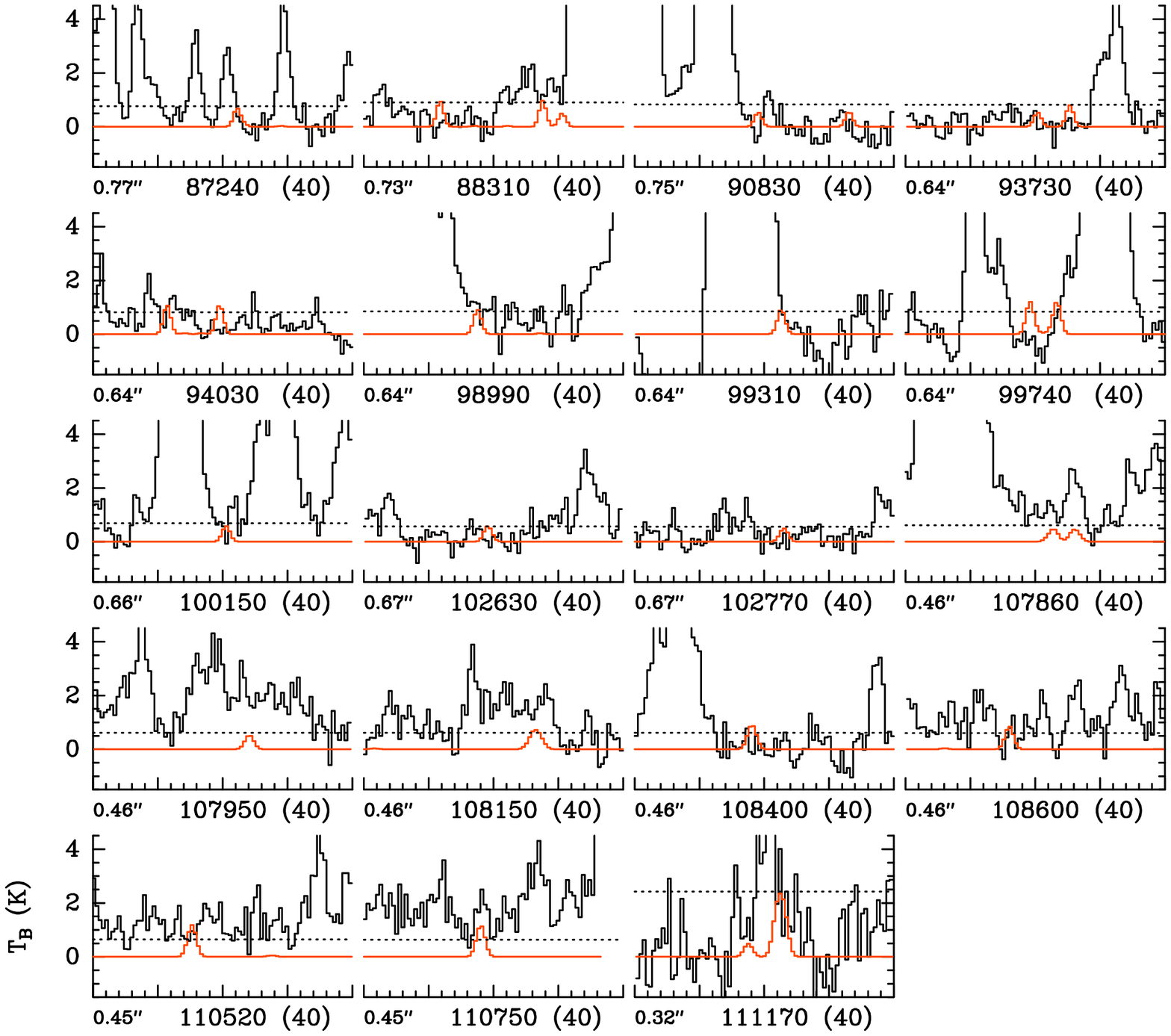}}}
\caption{Same as Fig.~\ref{f:spec_c2h5conh2_ve0_p158m272} but for propionamide,
C$_2$H$_5$C(O)NH$_2$, $\varv_{30}=1$.}
\label{f:spec_c2h5conh2_v30e1_p158m272}
\end{figure*}

\end{appendix}

\end{document}